%% file: tweb.tex
\pdfoutput=1
\documentclass[prodmode,acmtweb]{acmsmall} 
\usepackage{amsmath}
\usepackage{mathptmx}
\usepackage{amssymb}

\usepackage[utf8]{inputenc}
\usepackage[T1]{fontenc}
\usepackage{longtable}
\usepackage[normalem]{ulem}
\usepackage{multirow}
\usepackage{indentfirst}
\usepackage{graphicx}

\usepackage[ruled]{algorithm2e}

\SetAlFnt{\small}
\SetAlCapFnt{\small}
\SetAlCapNameFnt{\small}
\SetAlCapHSkip{0pt}
\IncMargin{-\parindent}

\begin{document}

\markboth{de Paula et al.}{Detecting and Handling Flash-Crowd Events on Cloud Environments}

\title{Detecting and Handling Flash-Crowd Events on Cloud Environments}
\author{Ubiratam~de~Paula
\affil{Fluminense Federal University}
Daniel~de~Oliveira
\affil{Fluminense Federal University}
Yuri~Frota
\affil{Fluminense Federal University}
Valmir~C.~Barbosa
\affil{Federal University of Rio de Janeiro}
 L\'{u}cia~Drummond
\affil{Fluminense Federal University}
\begin{center}
\textit{Submitted to the ACM Transactions on the Web (TWEB).}
\end{center}
}


\begin{abstract}
	Cloud computing is a highly scalable computing paradigm where resources are delivered to users on demand via Internet. There are several areas that can benefit from cloud computing and one in special is gaining much attention: the flash-crowd handling. Flash-crowd events happen when servers are unable to handle the volume of requests for a specific content (or a set of contents) that actually reach it, thus causing some requests to be denied. For the handling of flash-crowd events in Web applications, clouds can offer elastic computing and storage capacity during these events in order to process all requests. However, it is important that flash-crowd events are quickly detected and the amount of resources to be instantiated during flash crowds is correctly estimated. In this paper, a new mechanism for detection of flash crowds based on concepts of entropy and total correlation is proposed. Moreover, the Flash-Crowd Handling Problem (FCHP) is precisely defined and formulated as an integer programming problem. A new algorithm for solving it, named FCHP-ILS, is also proposed. With FCHP-ILS the Web provider is able to replicate contents in the available resources and define the types and amount of resources to instantiate in the cloud during a flash-crowd event. Finally we present a case study, based on a synthetic dataset representing flash-crowd events in small scenarios aiming at comparing the proposed approach with de facto standard Amazon’s Auto Scaling mechanism.
\end{abstract}

%
%


%
%

\terms{Flash Crowd, Cloud Computing, Optimization}


\acmformat{Ubiratam~de~Paula, Daniel~de~Oliveira, Yuri~Frota, Valmir~C.~Barbosa, L\'{u}cia~Drummond, 2015. Detecting and Handling Flash-Crowd Events on Cloud Environments.}

\begin{bottomstuff}
Author's addresses: U. de Paula, L. Drummond, D. de Oliveira, Y. Frota, Institute
of Computing, Fluminense Federal University, Niterói, Brasil; V. C. Barbosa, COPPE, Federal University of Rio de Janeiro, Rio de Janeiro, Brasil.
\end{bottomstuff}

\maketitle

\input{introduction}
\input{related_workf}
\input{detection}

\input{flashTraces}

\input{detection_results}
\input{definition}

\input{heuristic}

\input{results_FCHP}

\input{results_FCD_FCHP}
\input{conclusion}
\input{appendix}

\bibliographystyle{ACM-Reference-Format-Journals}
\bibliography{ref}

\end{document}

%% file: introduction.tex

\section{Introduction}
\label{sec:introduction}

	The emergence of high-speed networks and the popularization of mobile devices provoked the increase of the number of accesses to online content on the Internet \cite{stats}. This ever-growing number of potential users requires that Web servers are able to meet the Quality of Service (QoS) requirements such as availability and download speed. Currently, mainly because of the popularity of social networks (\textit{e.g.} Facebook, Twitter and Instagram), it is usual that a given content (e.g. a photo, a post or a video) has its popularity increased in a fast pace, becoming a viral. In many occasions, servers are unable to handle the volume of requests for the specific content (or a set of contents) that actually reaches it, thus causing some requests to be denied. This fact is named a flash-crowd event \cite{artigo-10}  and depending on its intensity can turn a content unavailable. For example, during the Oscars 2014 ceremony, Ellen DeGeneres twitted a selfie with several top movie celebrities \cite{selfie} . Ellen’s tweet reached incredible numbers (in a few hours) such as: (i) 8.1 million people saw the original tweet a total of 26 million times; (ii) 13,711 Web pages embedded the tweet, and those embeds were seen 6.8 million times; (iii) it has been seen 32.8 million times and (iv) it was retweeted more than 3.2 million times. This number of visualizations and accesses made Twitter page to be unable to be accessed by users for several minutes during Sunday night. In flash-crowd events such as the aforementioned one, all requests are legitimate, \textit{i.e.} the users really want to access the specific content. However, since the number of accesses increases in a short period of time, this can be a problem. Service providers may not be able to increase the number of Web servers to deploy these resources and the final effect is commonly a reduction in QoS. In more critical cases (such as 11th September attacks to World Trade Center), the continuous accesses to these Web servers can produce a complete (and undesired) halt.

	In order to avoid future problems in Web servers, flash-crowd events should be detected and treated as soon as possible. This treatment is usually accomplished by increasing the number of Web servers. Thus, clouds can be used to meet this emergency demand for computing resources \cite{Vaquero2008}. Cloud computing provides scalable computing resources to its users. These resources can be instantiated on demand, \textit{i.e.} users request the computing resources for a period of time and only pay what they use. Especially for handling of flash-crowd events in Web applications, clouds are suitable to provide more computing power and storage capacity during a specific time quantum. Thus, service providers can instantiate more resources during these events and then destroy these additional resources when flash-crowd events end. However, service providers have to estimate the amount of resources to instantiate to handle the flash-crowd event. The estimation has to be performed quickly and with no under or over estimations that could either slowdown the access to the content or produce high financial costs, respectively. In clouds, this estimation is a hard task to be accomplished due to the large number of heterogeneous resources provided by a cloud provider.

	Many papers in literature use only information about the number of content accesses to detect of flash-crowd events \cite{artigo-11}, \cite{artigo-1}. These papers try to quantify the increase in the number of accesses through certain characteristics, such as exponential growth \cite{artigo-11}. However, it is very difficult to define a scale for measuring the increase in the number of accesses. Furthermore, the increase in the number of accesses, even exponential, may not represent a flash-crowd event. This increase can be only a continued growth in content demand or a DDoS attack\cite{Monitoring}. Thus, it is top priority to correctly detect and handle flash-crowd events as soon as possible.

	In this paper, a new mechanism to detect flash crowds named Flash-Crowd Detection (FCD), which uses concepts of entropy and total correlation, is proposed. Additionally, the Flash-Crowd Handling Problem (FCHP) is precisely defined and formulated as an integer programming problem. A new algorithm for handling with flash crowds named FCHP-ILS is also proposed. FCHP-ILS is based on a metaheuristic called ILS (Iterated Local Search). With FCHP-ILS the service provider can define the types of resources (\textit{i.e.} Web servers) to instantiate in the cloud during a flash crowd. This estimation uses as input several types of resources, characterized by their total storage and bandwidth, available in the cloud and client requests. The FCHP-ILS was combined with the FCD mechanism in order to provide detection and handling in a single approach. The service provider can define the types of resources (\textit{i.e.} Web servers) to instantiate in the cloud during a flash crowd using FCHP-ILS when the flash crowd is detected by FCD. 

We compare the results of FCHP-ILS with the exact solution given by the proposed model, by using as input synthetic and real flash-crowd traces obtained from the related literature \cite{copa98}. We also present a comparison with the proposed approach with Amazon's Auto-Scaling mechanism \cite{AS}, the \textit{de facto} standard mechanism in the cloud. Both results showed that the proposed approach is capable to detect a flash-crowd event and precisely estimate the amount of resources to instantiate in the cloud environment, presenting some advantages when compared with the Auto Scaling mechanism of Amazon. 

%% file: related_workf.tex
\section{Related Work}
\label{rel_work}

In this section we describe some related work which propose solutions for handling flash-crowds. Since many of the proposed approaches either focus on detection or treatment we present 2 sub-sections, one for each topic.

\subsection{Detecting Flash Crowds}

	Several papers in the literature characterize flash-crowd events and use these characteristics in its detection \cite{artigo-11}, \cite{artigo-1}, \cite{artigo-2}. These papers commonly consider the beginning of a flash crowd when the increase in the number of accesses can be mapped as an exponential function or when they exceed defined thresholds.

	Wendell \textit{et al.} \cite{artigo-11} define flash crowd as a period over which request rates for a particular fully-qualified domain are increasing exponentially. Flash crowds are also defined by Stavrou \textit{et al.} \cite{artigo-1} as a phenomenon that results from a sudden, unpredicted increase in an on-line content's popularity. 

	Ari \textit{et al.} \cite{artigo-2} present a model of flash-crowd events, considering three major phases, a \textit{ramp-up phase}, a \textit{sustained traffic phase} and a \textit{ramp-down phase}. Until the \textit{ramp-up phase}, the traffic is considered \textit{normal}. In the \textit{ramp-up phase}, the traffic rises significantly for a small number of contents. After this rise, the traffic keeps high for some time (depending on the type of the flash crow), in the \textit{sustained traffic phase}. In the \textit{ramp-down phase}, the number of accesses to this small number of contents begin to fall.

	Similarly, Chen \textit{et al.} \cite{Chen} propose an adaptive admission control mechanism named NEWS that it is able to detect flash crowds based on performance degradations.

	Differently from the two previous papers, Sladescu \textit{et al.} \cite{Polymorphic} proposes a model to identify bursts, which are parameterized by a single set of burst feature types. In summary, Sladescu et al. propose a set of burst profiles that can be used to identify these events. The authors claim that these profiles can differ for different event types, and will depend on a variable number of feature types that describe the burst's associated event. The main drawback of this approach is that all proposed profiles are not able to characterize all types of flash-crowd events.
	
	Xie and Yu \cite{Monitoring} propose a novel detector of anomaly events based on hidden Markov model. This approach is based on the use of entropy of document popularity to detect the potential attacks. Although this approach is based on concepts used in this article, their focus is on DDoS attacks, which are a different scenario.
	
\subsection{Handling Flash Crowds}

	Several related work tackle the flash-crowd handling problem once the flash crowd was properly detected. A complete study about flash-crowd problem is provided in Jung \textit{et al.} \cite{artigo-10}. In this study, many solutions are focused on applications that use Content Distribution Networks (CDN), while some can also be applied generally to any type of Web applications. In addition, we note that the possibility of instantiating new resources on demand using clouds can also provide new solutions to handle with flash crowds. Broberg \textit{et al.} \cite{metacdn} propose a general-purpose framework called MetaCDN, which interacts with cloud providers to implement an overlay network that can be used as a CDN in the cloud.

	Other papers propose mechanisms to handle flash-crowd using P2P networks \cite{fcan2,fcan,artigo-1}. Pan and Atajanov \cite{fcan2,fcan} propose an approach named FCAN, that implements a P2P overlay over the real network to distribute the flash traffic from origin Web server. It is based on DNS redirection to route the requests in a balanced way; however, it does not increase or decrease the amount of resources during the flash-crowd. Stavrou \textit{et al.} \cite{artigo-1}  propose a system called PROOFS that implements a P2P overlay, which allows clients that seek popular contents to obtain them from other clients.

Some researches have also proposed frameworks to predict and to mitigate flash crowds through dynamic resource allocation \cite{Moore}, \cite{Vasar}, \cite{tang2014}.
Moore \textit{et al.} \cite{Moore} propose an elasticity management framework that consider as input a series of reactive rule-based strategies and generates a proactive strategy as outcome. They combine reactive and predictive auto-scaling techniques, \textit{i.e.} they try to predict when a flash crowd (they call as peaks) will occur.
Vasar \textit{et al.} \cite{Vasar} propose a framework that integrates a set of monitoring tools. The framework supports dynamic server allocation based on incoming load using a response-time-aware heuristic. 
Tang \textit{et al.} \cite{tang2014} propose a systematic framework for dynamic request allocation and service capacity scaling in a cloud-centric media network. 

There are also some commercial solutions such as the combination of Amazon's Auto Scaling and Load Balancing mechanisms \cite{AS,ELB}. Amazon's Auto Scaling mechanism allow for users to horizontally scale the amount of virtual machines according to current environment status, \textit{i.e.} during a flash crowd the number of virtual machines the user has can be increased to maintain performance, and decreases automatically during demand lulls to minimize financial costs. Flash crowds are identified in Amazon's environment using the cloud watch mechanism that allows to identify when the income traffic trespassed a pre-defined threshold. General Web applications are suitable to benefit from Auto Scaling mechanism. However, Auto Scaling only creates the virtual machines. To provide load balancing we have to use the Load Balancing mechanism. The Amazon's Elastic Load Balancing (ELB) automatically distributes incoming requests over the multiple instantiated virtual machines.
However, providing these kinds of scaling rules in commercial solutions is difficult, error-prone and asks some infrastructure expertise. In this way, Kouki and Dedoux \cite{SCAling} propose SCAling, a platform and an approach driven by Service Level Agreement (a formal contract between a service provider and a service consumer on an expected QoS level) requirements for Cloud auto-scaling.

Although there are many related papers in the literature, to the best of our knowledge none of them uses concepts of the information theory for detecting flash-crowds events or treats the problem considering different characteristics of the real problem jointly. 

%% file: detection.tex
\section{Flash-Crowd Detection Mechanism}
\label{sec:detection}

In this section we describe the concept of total correlation that is used in the proposed Flash-Crowd Detection mechanism, named FCD. 
The total correlation of a group of random variables, given a joint distribution, is a measure that quantifies the redundancy or dependence among them \cite{correlacao_total}.
This concept is based on the information-theoretic notion of entropy, itself a measure of uncertainty based on the same joint distribution, often referred to as the Shannon entropy \cite{shannon48}.
In the case of two variables, total correlation coincides with the variables' mutual information \cite{Han80a}.

In a flash-crowd event, the number of accesses to a certain set of contents increases significantly, while for the other contents the number of accesses remains at the same level.
The onset of this scenario has an important impact on the uncertainty related to which contents will be accessed next, and consequently on how these future accesses correlate with what is
happening presently. Once appropriate random variables are identified and their distributions quantified, the notion of total correlation can therefore be useful in
helping characterize the occurrence of a flash crowd.

Henceforth we assume that time is discrete and given by a nonnegative integer. We define two random variables, one relative to some time $t$, the other relative to time $t'=t-w$ for some integer $w>0$,
whence $t\geq w$.
The first random variable is denoted by $X$ and represents a content accessed at time $t'$. Similarly, random variable $Y$ is a content
accessed at time $t$. We assume that contents are identified by nonnegative integers as well, whose set is then the domain of both $X$ and $Y$.

By simply observing all accesses at times $t'$ and $t$, clearly the following probability mass functions ($f$ and $g$) and corresponding cumulative distributions ($F$ and $G$) can be easily estimated:
\begin{eqnarray*}
f(x)=\Pr(X=x);&F(x)=\Pr(X \leq x);\\
g(y)=\Pr(Y=y);&G(y)=\Pr(Y \leq y),
\end{eqnarray*}
where $x$ and $y$ identify specific contents.
Depending on the value of $w$, variables $X$ and $Y$ can be more or less correlated with each other. One way to measure this correlation in the absence of any further information on how the
two variables are jointly distributed is to calculate the Pearson sample correlation coefficient, normally denoted by $\rho$ and given by
\begin{equation}\label{rho}
\rho = \dfrac{\sum_{x\geq 0} (c'_x - a') (c_x - a)}{\sqrt{\sum_{x\geq 0} (c'_x - a')^2}  \sqrt{\sum_{x\geq 0} (c_x - a)^2}},
\end{equation}
where $x\in\{0,1,\ldots\}$ identifies the possible contents, $c'_x$ is the number of accesses to content $x$ at time $t'$, and $c_x$ is the number of accesses to $x$ at time $t$, with
averages $a'$ and $a$, respectively. In Equation \ref{rho}, $\rho$ is an estimate of the Pearson correlation coefficient between variables $X$ and $Y$, based on data alone.

Should the joint probability mass function of $X$ and $Y$, call it $p(x,y)=\Pr(X=x,Y=y)$, be also known [and consequently also the corresponding cumulative distribution, call it $P(x,y)=\Pr(X \leq x,Y \leq y)$],
it would be possible to calculate the Pearson correlation coefficient itself, since it is given by
\begin{equation}\label{rho'}
\hat{\rho}=\dfrac{E(XY)-E(X) E(Y)}{\sqrt{E(X^2)-(E(X))^2}  \sqrt{E(Y^2)-(E(Y))^2}},
\end{equation}	
where
\[
E(X) = \sum_{x\ge 0} x  f(x),
\]
\[
E(Y) = \sum_{y\ge 0} y  g(y),
\]
\[
E(X^2) = \sum_{x\ge 0} x^2  f(x),
\]
\[
E(Y^2) = \sum_{y\ge 0} y^2  g(y),
\] 
and
\[
E(XY) = \sum_{x\ge 0}\sum_{y\ge 0} x  y  p(x,y).
\]
As we show next, knowing $p(x,y)$ is also essential for the total correlation between $X$ and $Y$ to be calculated. This mass function is unknown, but adopting the value of $\rho$ as the target value
of (the also unknown) $\hat{\rho}$ allows $p(x,y)$ to be obtained in such a way as to be consistent with both $f(x)$ and $g(y)$.\footnote{Note that, although $f(x)$ and $g(y)$ follow from $p(x,y)$ via
$f(x)=\sum_{y\geq 0}p(x,y)$ and $g(y)=\sum_{x\geq 0}p(x,y)$, knowing $f(x)$ and $g(y)$ does not in general imply knowing $p(x,y)$ as well.}

The total correlation $C(X,Y)$ between $X$ and $Y$ is defined as
\begin{equation}\label{correlation}
C(X,Y)=H(X)+H(Y)-H(X,Y),
\end{equation}	
where $H(X)$ and $H(Y)$ are (marginal) entropies, given respectively by
\begin{equation}\label{H(X)}
\displaystyle H(X)=-\sum_{x\geq0} f(x) \log_2  f(x) 
\end{equation}
and
\begin{equation}\label{H(Y)}
\displaystyle H(Y)=-\sum_{y\geq0} g(y) \log_2  g(y),
\end{equation}
and $H(X,Y)$ is the (joint) entropy
\begin{equation}\label{H(X,Y)}
\displaystyle H(X,Y)=-\sum_{x\geq0} \sum_{y\geq0} p(x,y) \log_2  p(x,y).
\end{equation}
The easiest case is that of independent random variables, that is, the case in which $p(x,y)=f(x)g(y)$ for all $x,y\geq 0$. In this case, $H(X,Y)=H(X)+H(Y)$ and $C(X,Y)=0$.
All other cases require further work for the determination of $p(x,y)$, as we discuss next, following \cite{nelsen1987}.

We begin by writing two auxiliary joint cumulative distributions, both based on the already known $F(x)$ and $G(y)$:
\begin{equation}\label{H_L}
P_L(x,y)=\max\{F(x)+G(y)-1,0\};
\end{equation}	
\begin{equation}\label{H_U}
P_U(x,y)=\min\{F(x),G(y)\}.
\end{equation}	
From these, computing the corresponding mass functions, namely $p_L(x,y)$ and $p_U(x,y)$, is a simple matter, since
\begin{equation}\label{h_L}
p_L(x,y)=P_L(x,y)-P_L(x-1,y)-P_L(x,y-1)+P_L(x-1,y-1)
\end{equation}	
and
\begin{equation}\label{h_U}
p_U(x,y)=P_U(x,y)-P_U(x-1,y)-P_U(x,y-1)+P_U(x-1,y-1).
\end{equation}
In these expressions, $P_L(u,v)=0$ whenever $u<0$ or $v<0$, and likewise for $P_U(u,v)$.

If $\rho_L$ and $\rho_U$ are the Pearson correlation coefficients resulting from Equation \ref{rho'} with $p_L(x,y)$ and $p_U(x,y)$ substituting for $p(x,y)$, respectively, then the well-known
discrete Fréchet bounds result in
\begin{equation}
P_L(x,y)\leq P(x,y)\leq P_U(x,y)
\end{equation}
for all $x,y\geq 0$, and in
\begin{equation}
\rho_L\leq \hat{\rho}\leq\rho_U.
\end{equation}
It follows from these inequalities that, in order for $\hat{\rho}$ [which is based on $p(x,y)$] to be equal to $\rho$ (which is based on the data), it suffices for $p(x,y)$ to be a $\rho$-dependent convex
combination of $f(x)g(y)$ (the joint mass function for the independent-variable case) and either $p_L(x,y)$ or $p_U(x,y)$. Specifically, for $\rho\leq 0$ we let $\theta=\rho/\rho_L$ and get
\begin{equation}\label{h1}
p(x,y)=\theta p_L(x,y)+(1-\theta)f(x)g(y).
\end{equation}
For $\rho\geq 0$, in turn, we let $\theta=\rho/\rho_U$ and get
\begin{equation}\label{h2}
p(x,y)=(1-\theta)f(x)g(y)+\theta p_U(x,y).
\end{equation}
Readily, the case in which $X$ and $Y$ are independent implies $\rho=0$ and consequently $p(x,y)=f(x)g(y)$, by either expression, as expected.

Algorithm \ref{alg_fdm} summarizes the calculation of $C(X,Y)$, the total correlation between variables $X$ and $Y$.

\begin{algorithm}[thp]
\caption{\footnotesize FCD \label{alg_fdm}}
	\For{$t:=1$ \textbf{to} $T$}{
	
		\ForAll{ objects accessed in times $t$ and $t'=t-w$}{
			Obtain the probability mass functions $f(x)$ and $g(y)$\;
			Obtain the corresponding cumulative distributions $F(x)$ and $G(y)$\;
		}
		Calculate $\rho$, using Equation \ref{rho}\;
		Obtain $P_L(x,y)$ or $P_U(x,y)$, depending on whether $\rho\leq 0$ or $\rho\geq 0$, respectively using Equation \ref{H_L} or \ref{H_U}, respectively\;
		Obtain the corresponding $p_L(x,y)$ or $p_U(x,y)$, using Equation \ref{h_L} or \ref{h_U}, respectively\;
		Obtain the corresponding $\rho_L$ or $\rho_U$, using Equation \ref{rho'}\;
		Obtain $p(x,y)$, using Equation \ref{h1} or \ref{h2}, again depending on the value of $\rho$\;
		Calculate $H(X)$, $H(Y)$, and $H(X,Y)$, using Equations \ref{H(X)}, \ref{H(Y)}, and \ref{H(X,Y)}, respectively\;
		Calculate $C(X,Y)$, using Equation \ref{correlation}\;
	}
\end{algorithm}

\begin{figure}[ht]
\centering
\begin{center}
\begin{tabular}{c c}
\includegraphics[scale=0.4]{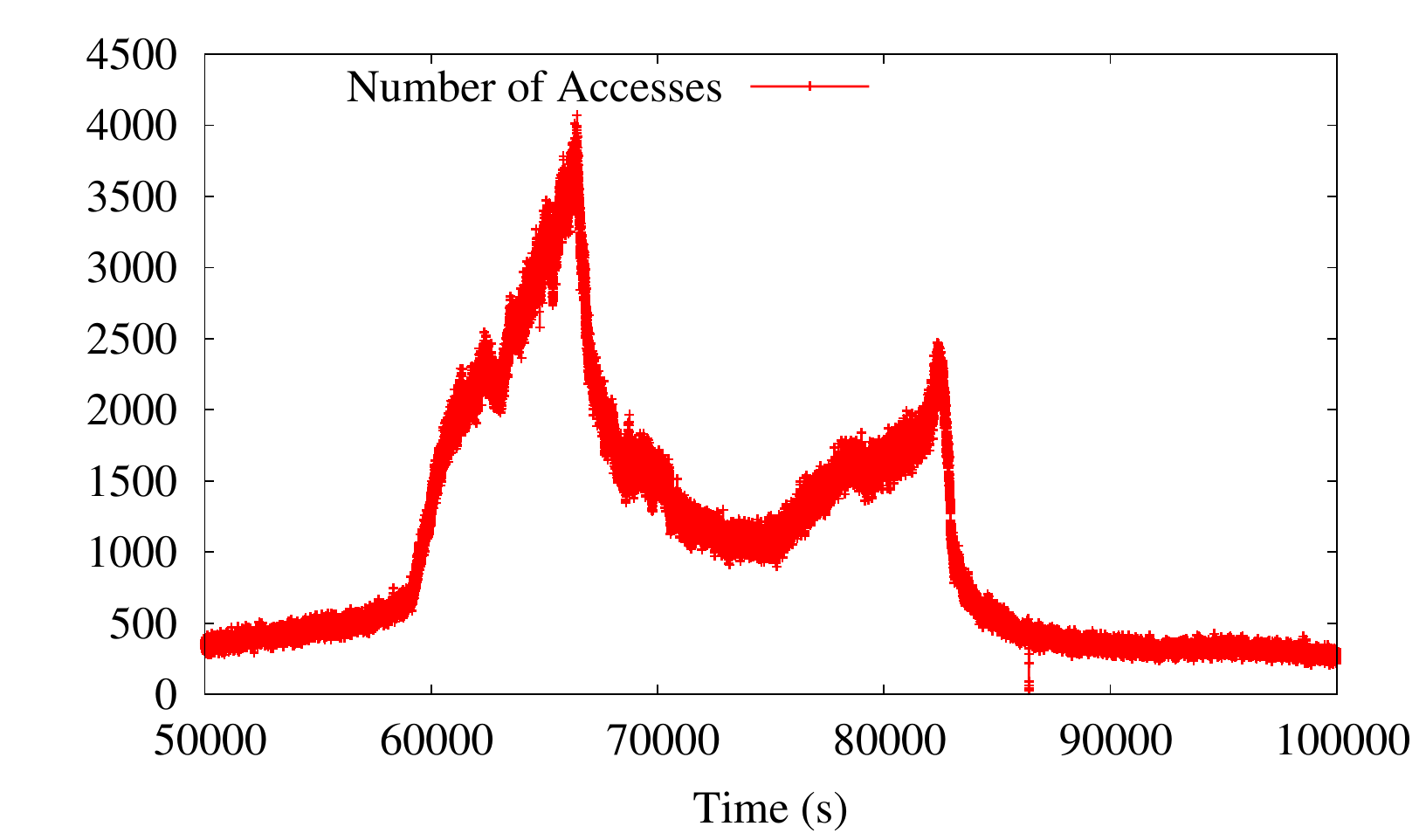} & \includegraphics[scale=0.4]{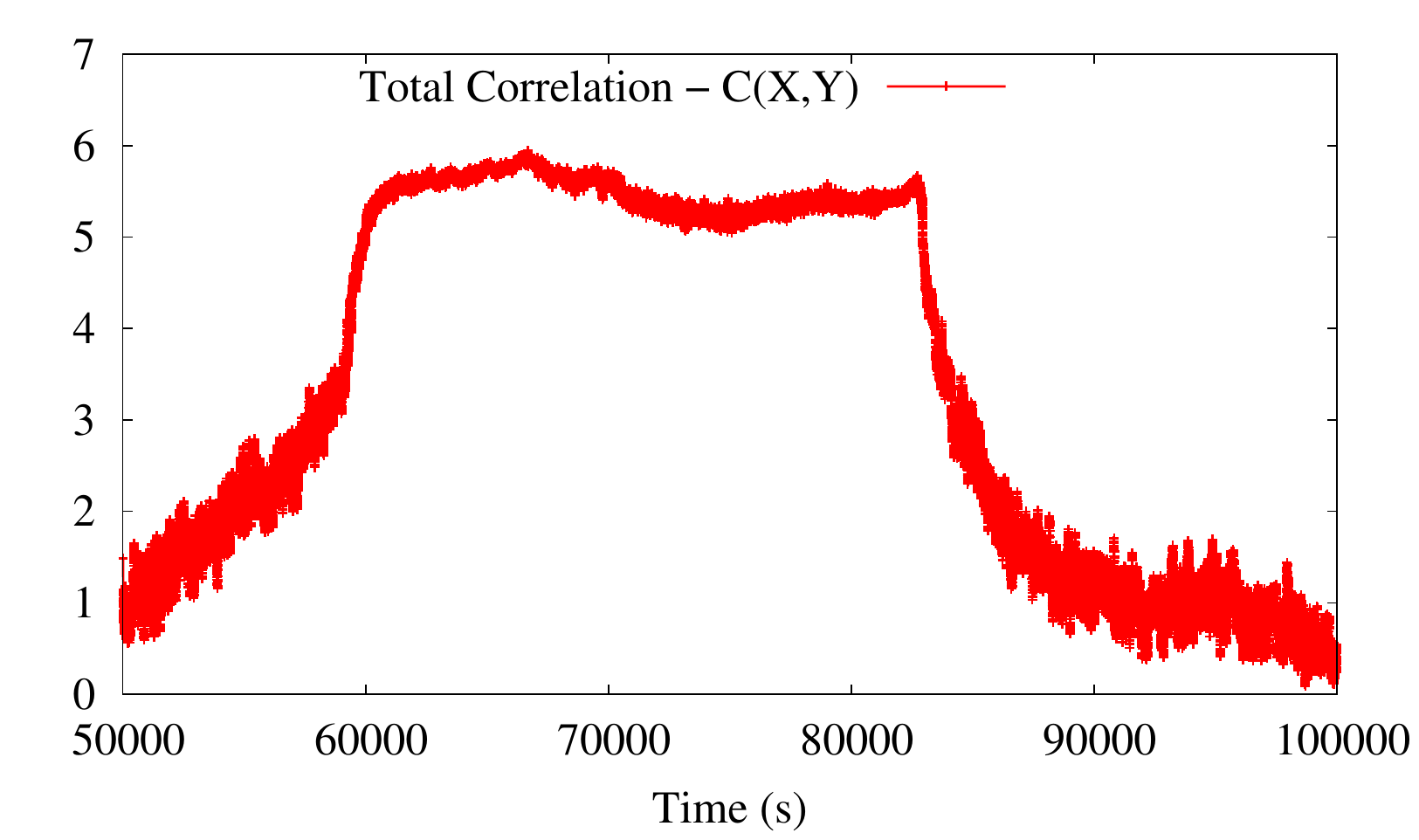} \\
(a) & (b) \\
\end{tabular}
\end{center}
\caption{Number of accesses (a) and total correlation for $w=1$ (b) on the 65$^{th}$ and $66^{th}$ days of the 1998 World Cup.}
\label{copa}
\end{figure}

The proposed flash-crowd detection mechanism uses total correlation, calculated as described, to analyze the accesses at each time $t\geq w$.
The motivating intuition is that, as a flash crowd begins and with it the uncertainty regarding which contents are accessed decreases (first at time $t$, then at both $t'$ and $t$ as $t$ progresses into the
flash crowd), all three entropies [$H(X)$, $H(Y)$, and $H(X,Y)$] have their values reduced. This reduction, however, is expected to be much more pronounced for $H(X,Y)$ than for either of the marginal entropies,
since the flash crowd acts to confirm the preeminent access to a same content at both time $t'$ and time $t$. That is, by elevating $p(u,v)$ or $p(v,u)$ for every content $u$ involved in the flash crowd
regardless of $v$, the joint entropy is expected to decrease by much more than the marginal ones. This, naturally, points at substantial increases in the total correlation $C(X,Y)$ as telltale signs that a
flash crowd is beginning. This is exactly what we see in the example given in Figure \ref{copa} for $w=1$, obtained from the access logs of the 65th and 66th days of the 1998 World
Cup.\footnote{\url{http://ita.ee.lbl.gov/html/contrib/WorldCup.html}} These data reveal a nontrivial flash crowd, beginning approximately at 60,000 seconds into the period depicted and ending at about
86,000 seconds. Both beginning and end are marked by substantial changes in total correlation, first up then down.

The main advantage of this approach is that it does away with the need to assume any functional dependence of the number of accesses on time (\textit{e.g.}, exponential). Instead, it uses entropy and total
correlation, which are bounded from below by $0$ and from above by functions of how many distinct contents there are, independently of how often they are accessed. If $n$ is this number of contents, then
the entropies $H(X)$ and $H(Y)$ are maximized by letting $f(x)$ and $g(y)$, respectively, be uniform, \textit{i.e.}, equal to $1/n$ for any content. It follows that
$H(X),H(Y)\leq -\sum_{x\geq 0} (1/n) \log_2(1/n) = \log_2 n$. Similarly, $H(X,Y)\leq -\sum_{x\geq 0}\sum_{y\geq 0} (1/n^2) \log_2(1/n^2) = 2\log_2 n$.
Finally, $C(X,Y)\leq 2\log_2 n$ as well (though equality can never be attained).

Once a flash-crowd event is detected in a Web application, it is necessary to mitigate its effects, such as unavailability and high latency in content access.

%% file: flashTraces.tex
\section{Flash-Crowd Traces}\label{sec:fctraces}

	In order to  evaluate the proposed  flash-crowd detection procedure in different scenarios, several kinds of traces were analyzed.   These traces can be classified as synthetic  or real world traces. The first category includes  traces created by a generator of synthetic flash crowds, presented in our previous work \cite{SAC2015}, and  a trace generated by the CloudSuite Web Serving benchmark. The second category comprises  real traces of the 1998 World Cup Web site \cite{copa98} and  of the idUFF system, collected by the Google Analytics service.
	

\subsection{Generator of Synthetic Flash Crowds}\label{generator}

	Due to the difficulty in obtaining real traces to evaluate the procedures for  detecting and handling flash crowds, a generator of synthetic flash crowds was developed. This generator can produce several kinds of traces by modeling flash crowd's main phases and taking into account some of the characteristics of real flash crowds. It is based on sampling from a variety of beta distributions with different parameters. Such  parameterization is convenient for  generation of flash crowd accesses with different shapes, particularly during the ramp-up and ramp-down phases, as  explained in detail in \cite{SAC2015}.

\subsection{CloudSuite Traces}

	CloudSuite is a benchmark suite for emerging scale-out applications, which consists of several applications selected due to  their popularity in today's data centers \cite{cloudsuite}.  One of these applications, a Web Serving benchmark, called CloudStone, consists of three main components: a Web server, a database backend and a client to emulate real world accesses to the Web server.  
	
	Although  CloudStone can be used to benchmark various Web technologies and Web server software stacks, in our work we are  interested  only in   the  workload composed by  the users accesses  generated by it. To create the synthetic  trace from  these accesses,  some extra  code was added in the CloudStone application. Thus,  we  were able to obtain  for each  content, the number of users accesses and the time when each access occurred.

\subsection{1998 World Cup Traces}

	This trace consists of all the requests made to the 1998 World Cup Web site between April 30, 1998 and July 26, 1998 \cite{copa98}.
Football fans around the world could access the current scores of the football matches in real time,  previous match results, player statistics and biographies, team histories, information about stadiums,
facts about local attractions and festivities, as well as a wide range of photos and sound clips from the matches and interviews with players and coaches.
	
	 During this period, all of the 33 different servers located at four geographic locations recorded time stamps with one second resolution. The access logs have been split into intervals of one day size. An example of a access log entry is:
	 \begin{verse}
		282 - - [30/Apr/1998:21:31:12 +0000] "GET /images/hm\_bg.jpg HTTP/1.0" 200 24736	 	
	 \end{verse}
	
	This entry shows that on time stamp \textit{[30/Apr/1998:21:31:12 +0000]}, the client \textit{282} asked for the file \textit{/images/hm\_bg.jpg} and reported that it supports \textit{HTTP/1.0}. Moreover, the server successfully responded to this request, indicated by status code of \textit{200}, and transferred \textit{24,736} bytes of content data to the client. Thus, each user's request recorded in the trace contains the following information: (i) \textit{clientID} (a unique identifier for the client), (ii) \textit{timestamp} (the time of the request), (iii) \textit{method} (the method contained in the client's request), (iv) \textit{objectID} (a unique identifier  for the requested URL), (v) \textit{object type} (the type of file requested generally based on the file extension), (vi) \textit{status} (HTTP version indicated in the client's request and response status code), and (vii) \textit{size} (the number of bytes in the response).
	 	 
	 
	 To detect a flash-crowd event only  the \textit{timestamp} and the \textit{objectID} information are necessary. They are used to calculate the total correlation for each interval of  time.	
	 The match days present  a flash-crowd event.  On days  with two matches,  two peaks of accesses happened. On the others days no flash-crowd events occurred.  
	
\subsection{Traces of idUFF}

	The idUFF System  from  Universidade Federal Fluminense (UFF)  aims at centralizing data of  students, professors and employees of the university.  At each beginning of a term, all  students from UFF must register for courses  they will attend in the semester, by using idUFF System.  So, during these periods, the idUFF's accesses increase significantly and, consequently, sometimes the system halts or becomes very slow.
	
	By using   a  traffic analysis tool for Web sites, it is possible to monitor the idUFF system  and   obtain information about all Web pages accessed during  certain periods of  time, such as  years, months, days, hours, weeks or minutes. 
Google Analytics service \cite{googleservice} is  commonly used for traffic analysis, giving detailed statistics   about the traffic of a Web site, like  the number of accesses for page, each one representing one content, and the time stamp, representing the period of time when each of these accesses occurred.  
	
	For our tests,  each record of the trace generated from the Google Analytics contains:  (i)  a page unique identification (\textit{e.g.} \textit{url}), (ii)  a time stamp  and (iii) the number   of accesses for such  page  associated with   the time stamp. Since we are interested in  flash-crowd events, we opted for  generating a  trace file  containing accesses accomplished during one enrollment day  and  with  each time stamp representing a period of sixty seconds.

%% file: detection_results.tex
\section{Evaluating the FCD Mechanism}

 The FCD mechanism  is  tested and evaluated over  each previously presented trace. For all tests, it was used $w=1$ to define the time window size applied in the detection mechanism. 

\subsection{Synthetic Flash Crowds}

	In this evaluation,  three synthetics traces, with different  number of contents and accesses,  and consequently representing  different flash crowds scenarios, were analyzed. 
	
	More specifically, in the first trace,  ten contents from a total of forty take part  of  a  flash-crowd event. The total number of accesses is 679,744, where 543,072 are made for flash-crowd contents. In the second one, twenty  from thirty contents  are involved in the flash-crowd with 3,540,739 accesses from a total of 3,645,927 accesses. Finally, the last trace has one hundred and twenty contents, but only  twenty of them  participate of the flash-crowd event. This trace contains an amount of 4,597,005 accesses and  3,540,739 of them are performed during the flash-crowd event.
	
\begin{figure}[!ht]
\centering
\begin{center}
\begin{tabular}{c c}
\includegraphics[scale=0.4]{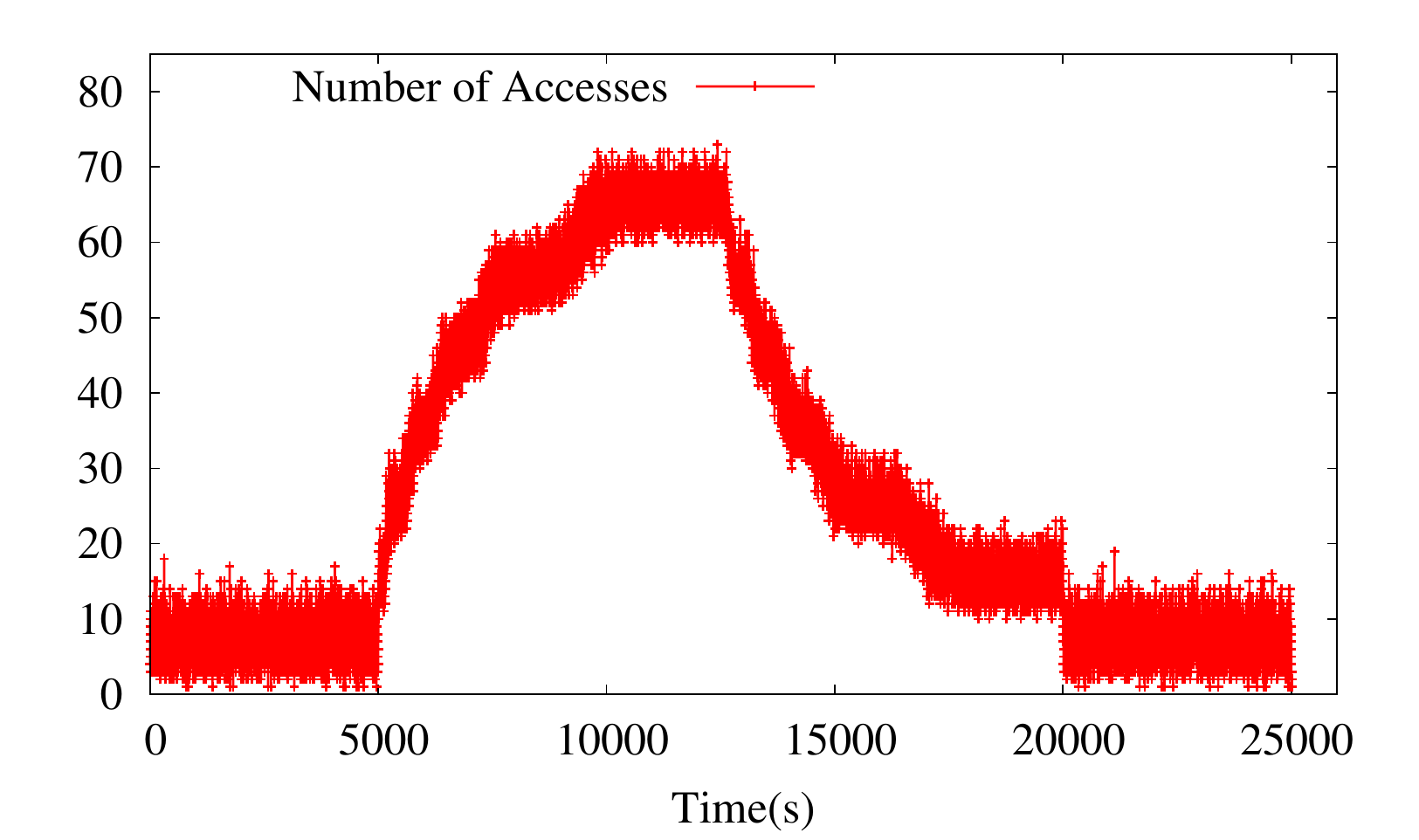} & \includegraphics[scale=0.4]{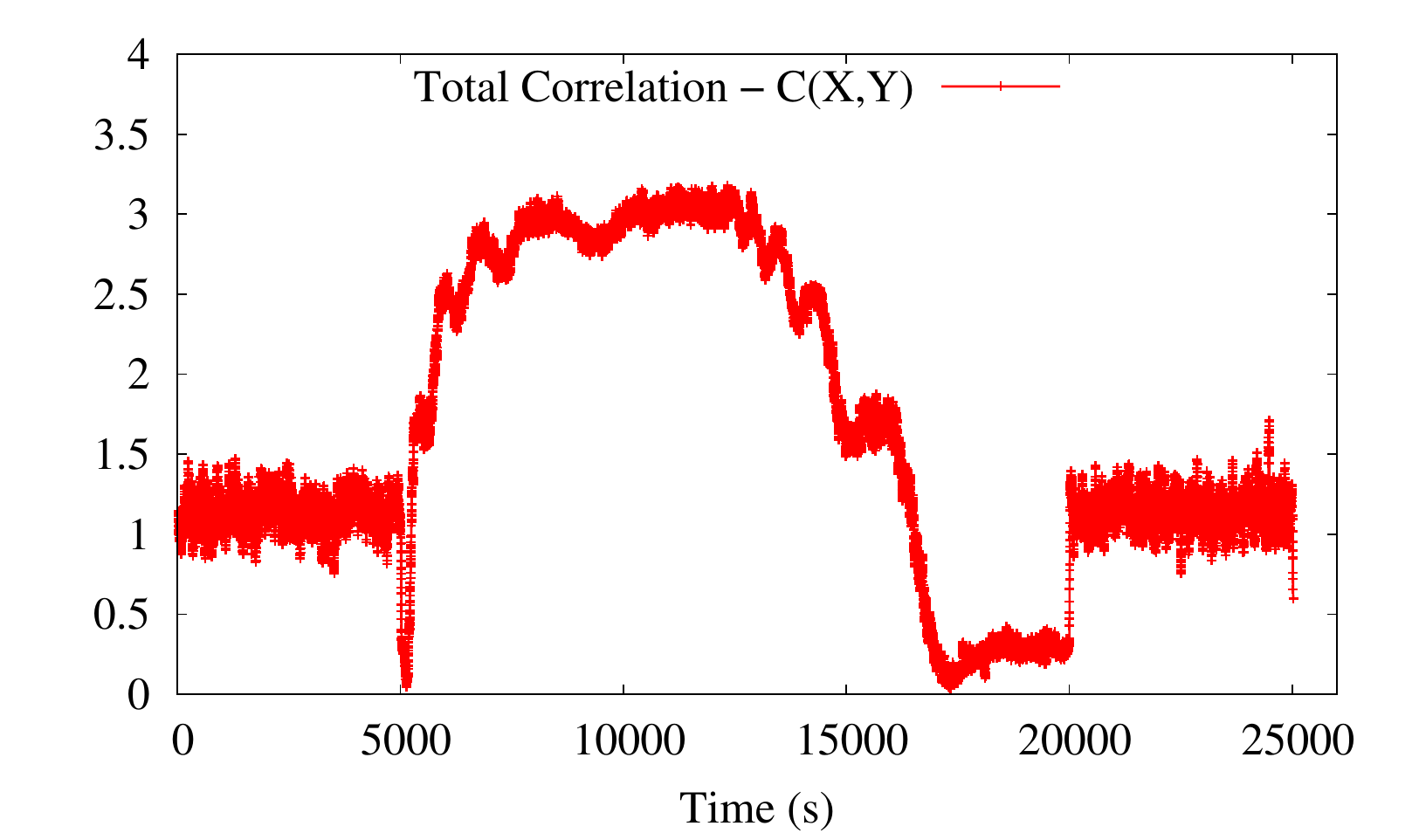} \\
(a) & (b) \\
\includegraphics[scale=0.4]{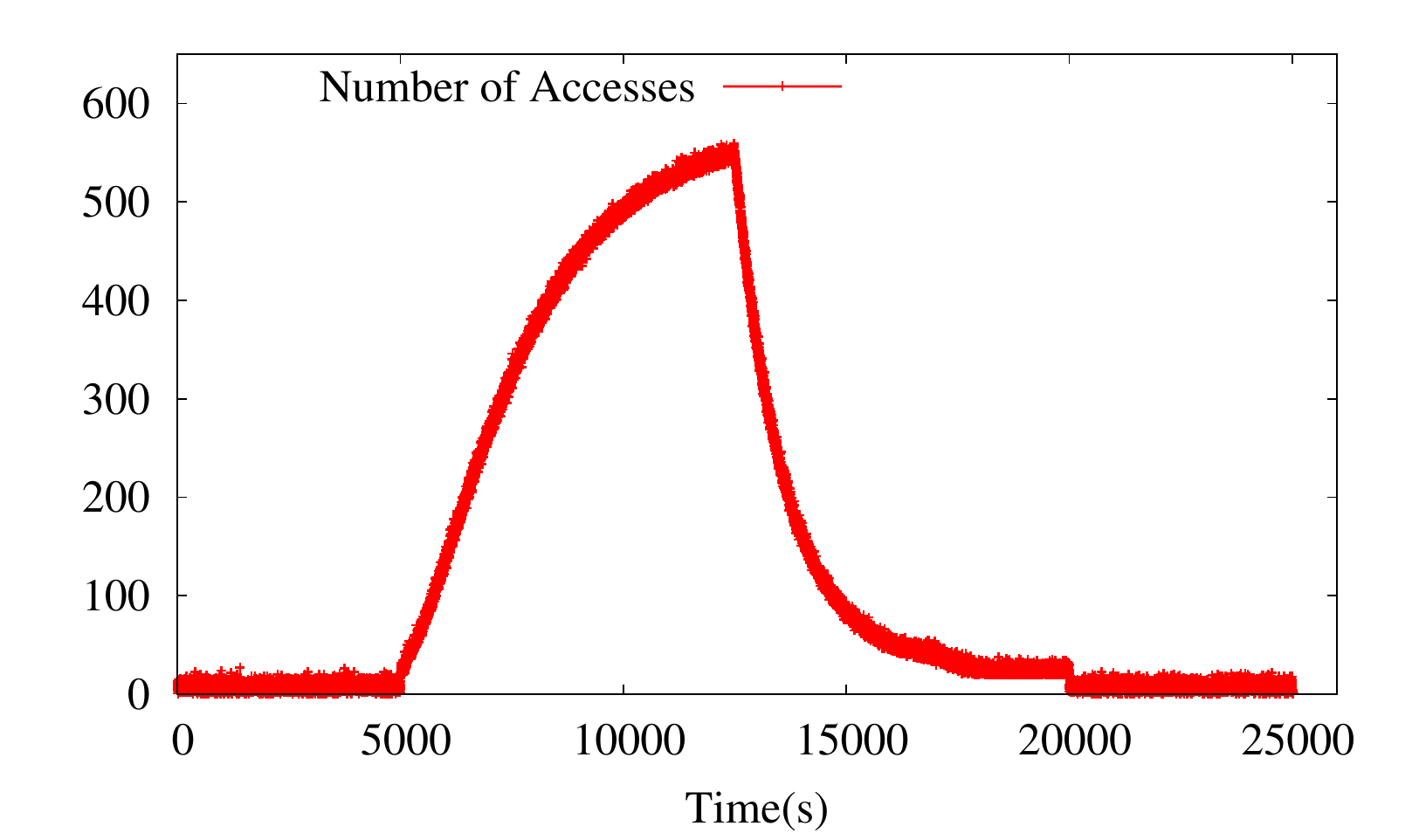} & \includegraphics[scale=0.4]{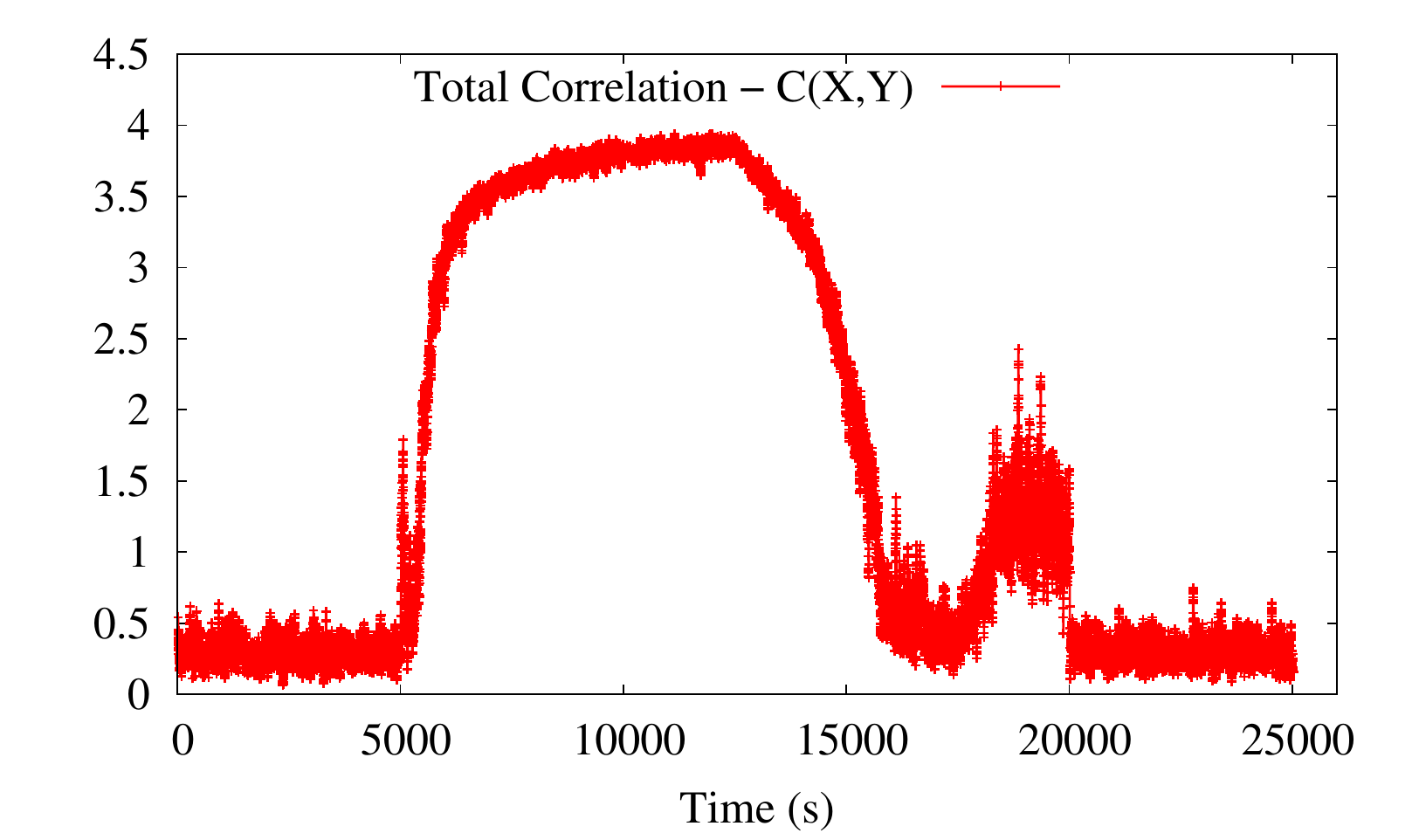} \\
(c) & (d)\\
\includegraphics[scale=0.4]{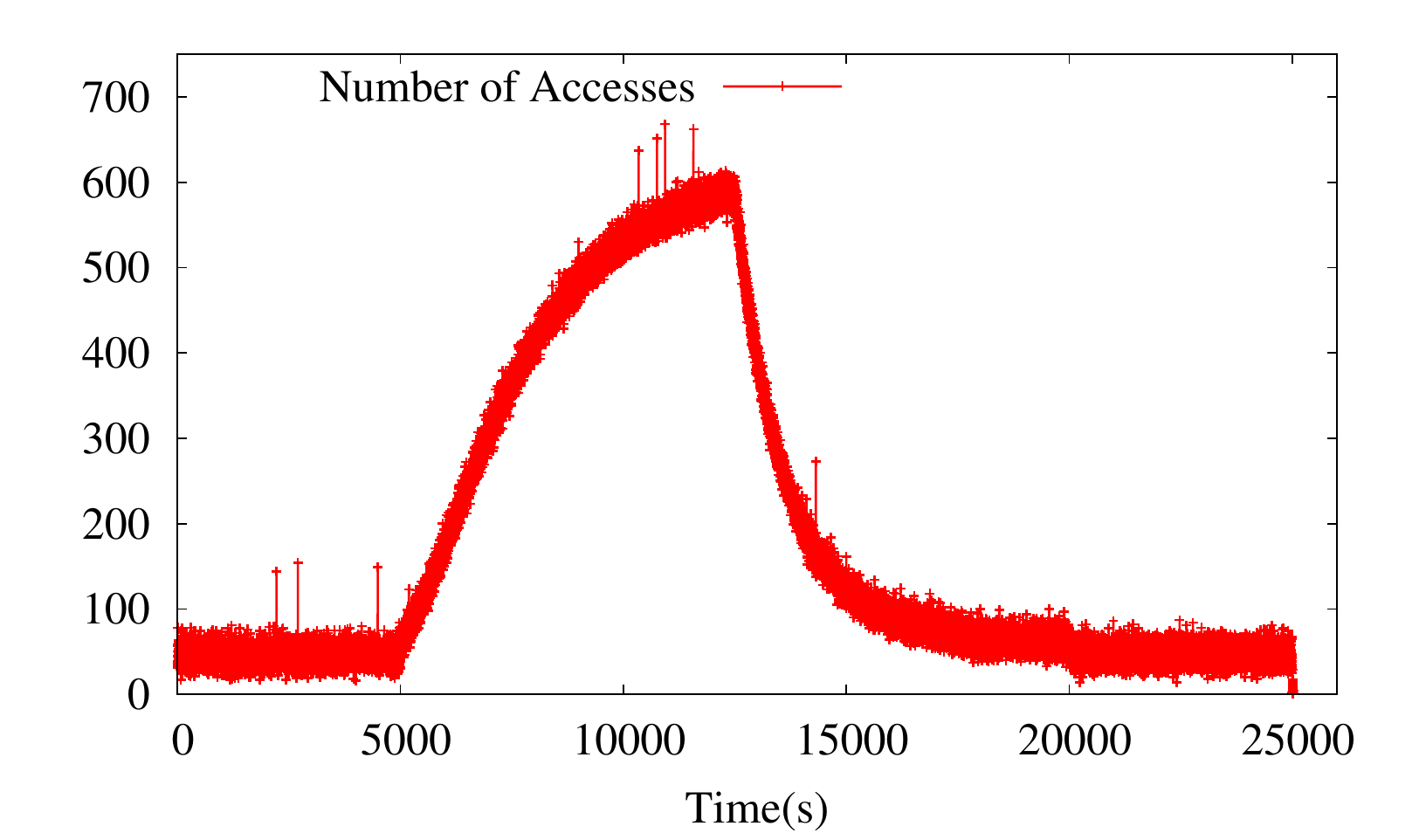} & \includegraphics[scale=0.4]{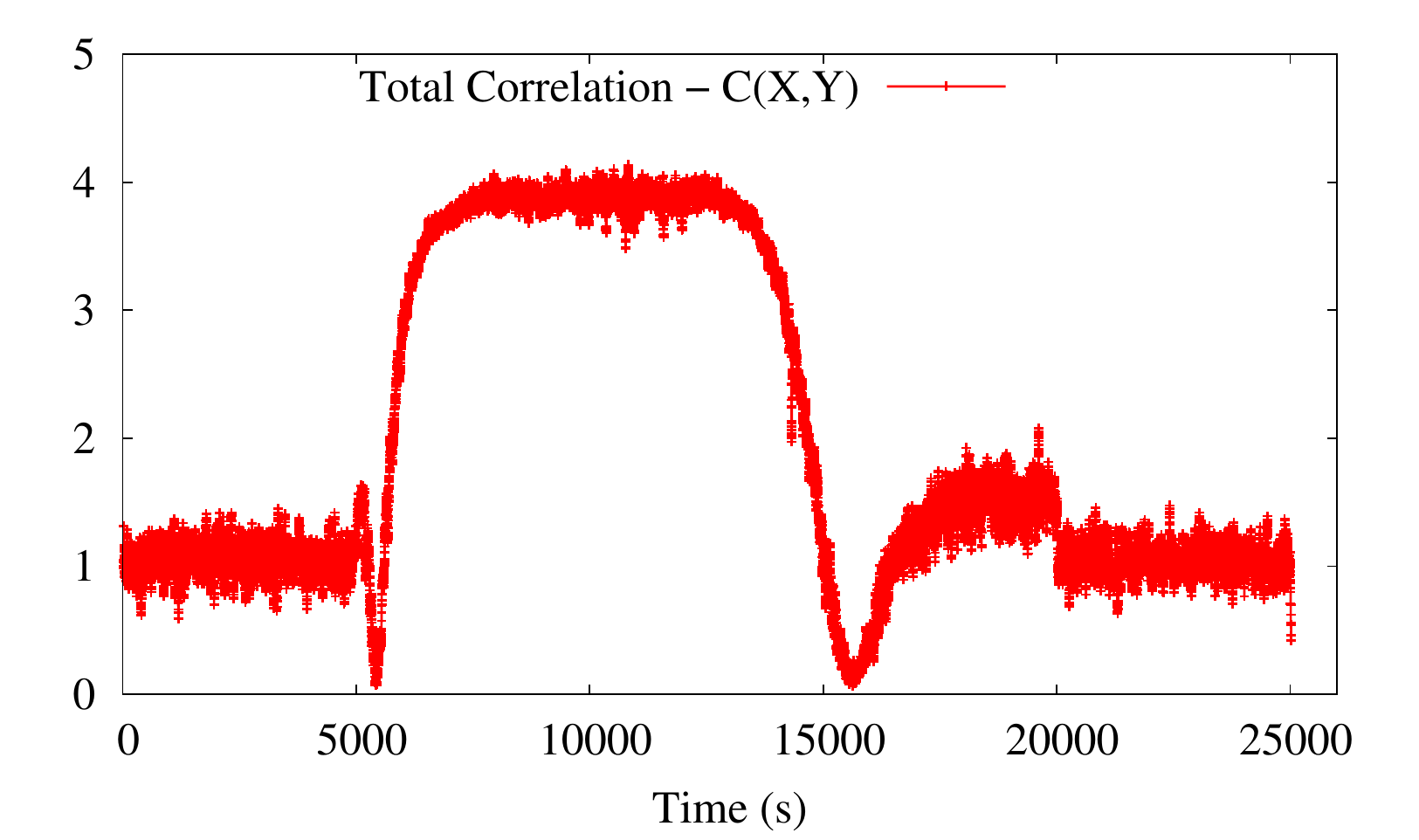} \\
(e) & (f)
\end{tabular}
\end{center}
\caption{Number of accesses (a),(c),(e) and total correlation for $w=1$ (b),(d),(f) for the synthetic traces with 40, 30 and 120 objects.}
\label{fc-syn-all}
\end{figure}

	Figures \ref{fc-syn-all}(a), \ref{fc-syn-all}(c) and \ref{fc-syn-all}(e) show the number of accesses for the three evaluated traces along the time.  Note that the flash-crowd event started approximately at 5,000 seconds and ended approximately at 15,000 seconds.
	
	Figures \ref{fc-syn-all}(b), \ref{fc-syn-all}(d), and \ref{fc-syn-all}(f) present the values of  the total correlation for all traces along the time. Independently of the scale of the number of  total contents, flash-crowd contents and  accesses (tens or hundreds of accesses per second), the FCD mechanism  correctly detected the beginning and the end of all flash-crowd events marked by substantial changes in the total correlation, at approximately 5,000 and 15,000 seconds.
	
%
%

\subsection{CloudSuite}
	The FCD mechanism was also evaluated over a synthetic trace generated by the CloudSuite. In this trace, ten from twenty contents are involved in a flash-crowd event. The total number of accesses is 516,435, where 485,482 are made for flash-crowd contents.

\begin{figure}[!ht]
\centering
\begin{center}
\begin{tabular}{c c}
\includegraphics[scale=0.4]{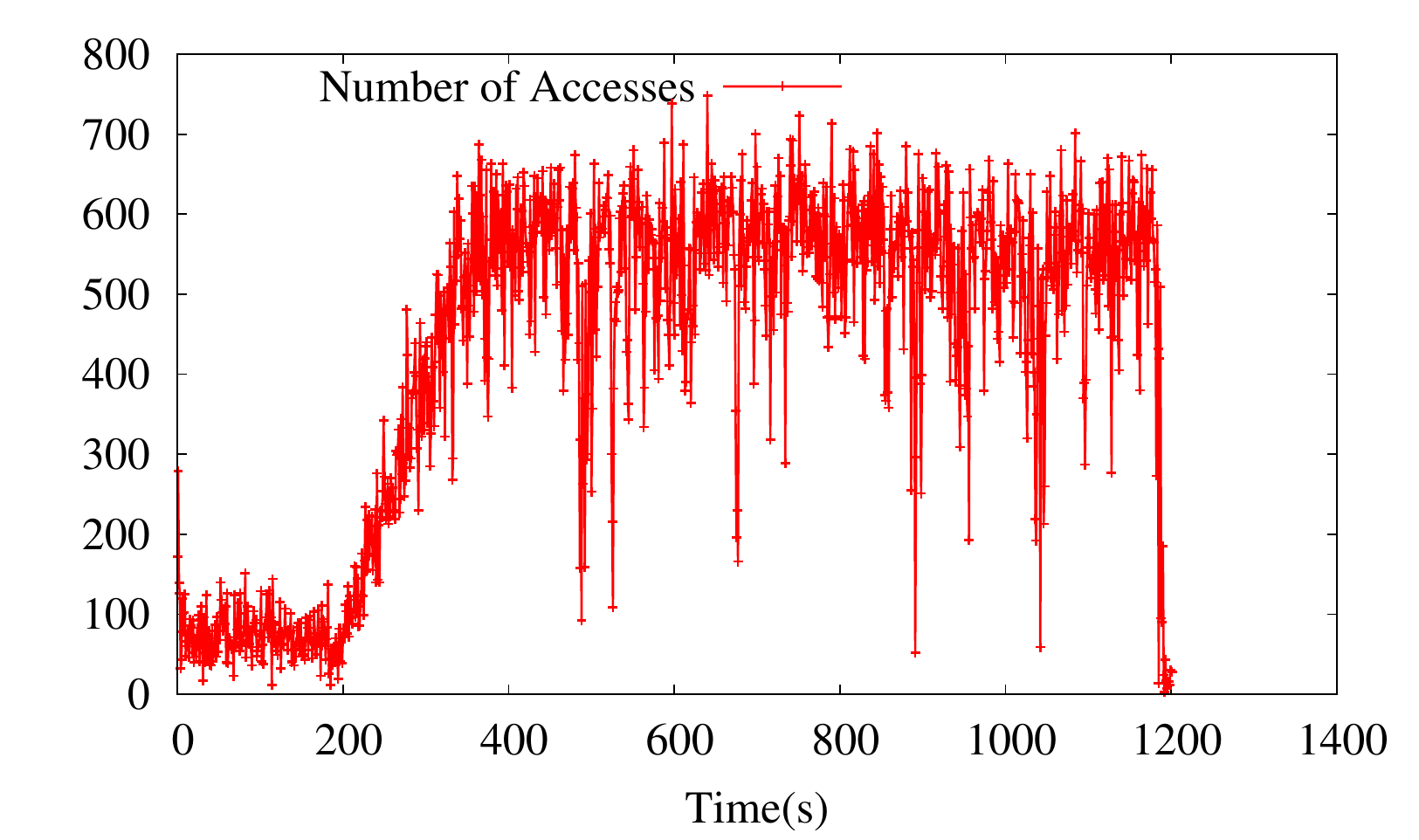} & \includegraphics[scale=0.4]{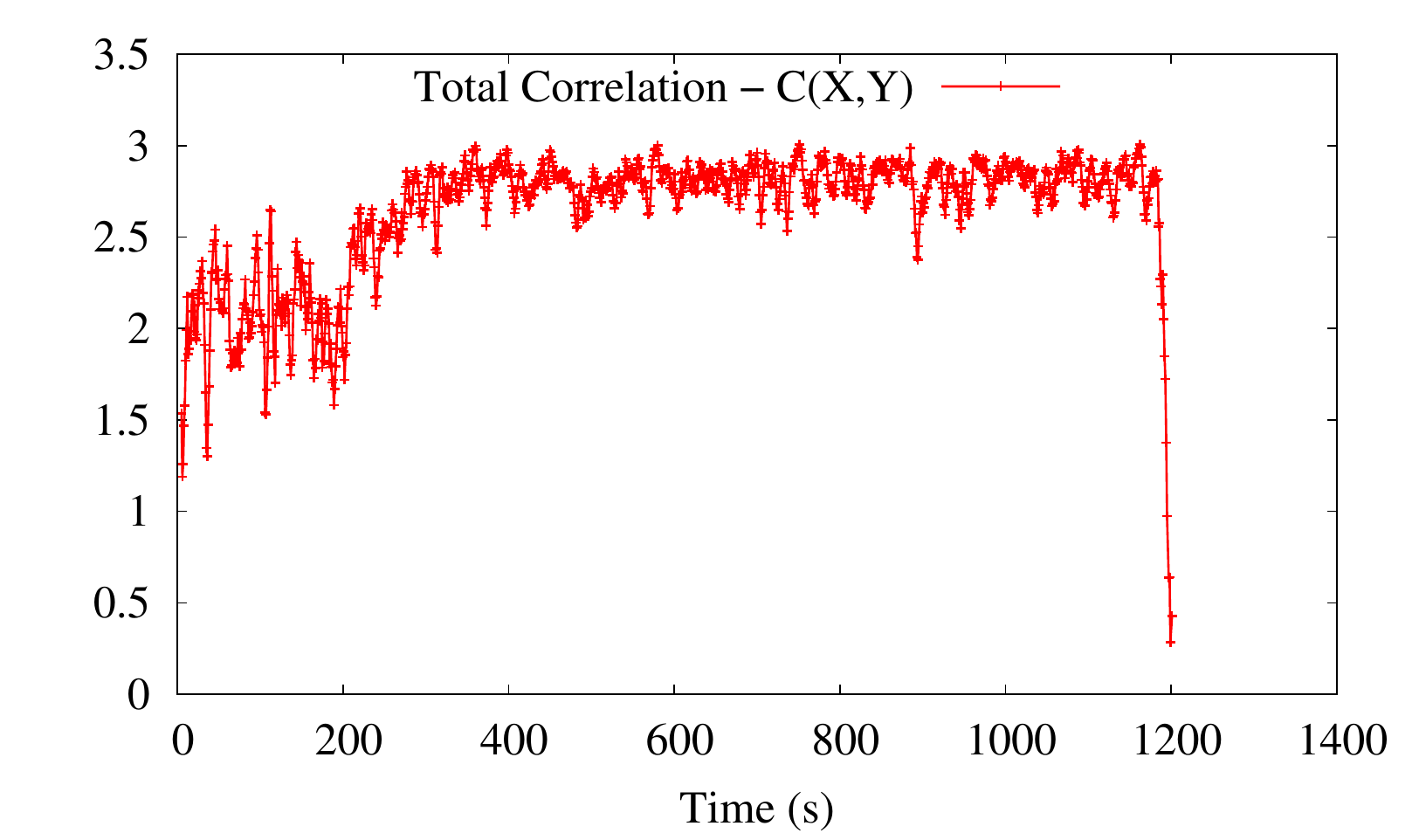} \\
(a) & (b) 
\end{tabular}
\end{center}
\caption{Number of accesses (a) and total correlation for $w=1$ (b) for the CloudSuite traces.}
\label{fc-cloudsuite}
\end{figure}

	Figure \ref{fc-cloudsuite}(a) shows the number of accesses along the time for the CloudSuite trace. The flash crowd started approximately at 200 seconds.
 As	Figure \ref{fc-cloudsuite}(b) shows, the total correlation also increases at this same time, indicating the FCD mechanism correctly detected the flash-crowd event.
%
%
%
%
	
\subsection{1998 World Cup}
	
	 In these tests, two traces of different days were used.     The first trace is formed by  accesses from  the  middle of the  66th and   beginning of 67th day of the World Cup, when two matches occurred and consequently two peaks of accesses happened.  The second one  starts during the 73th day  and  finishes in the 74th day, when only one match happened. 
	 
\begin{figure}[!ht]
\centering
\begin{center}
\begin{tabular}{c c}
\includegraphics[scale=0.4]{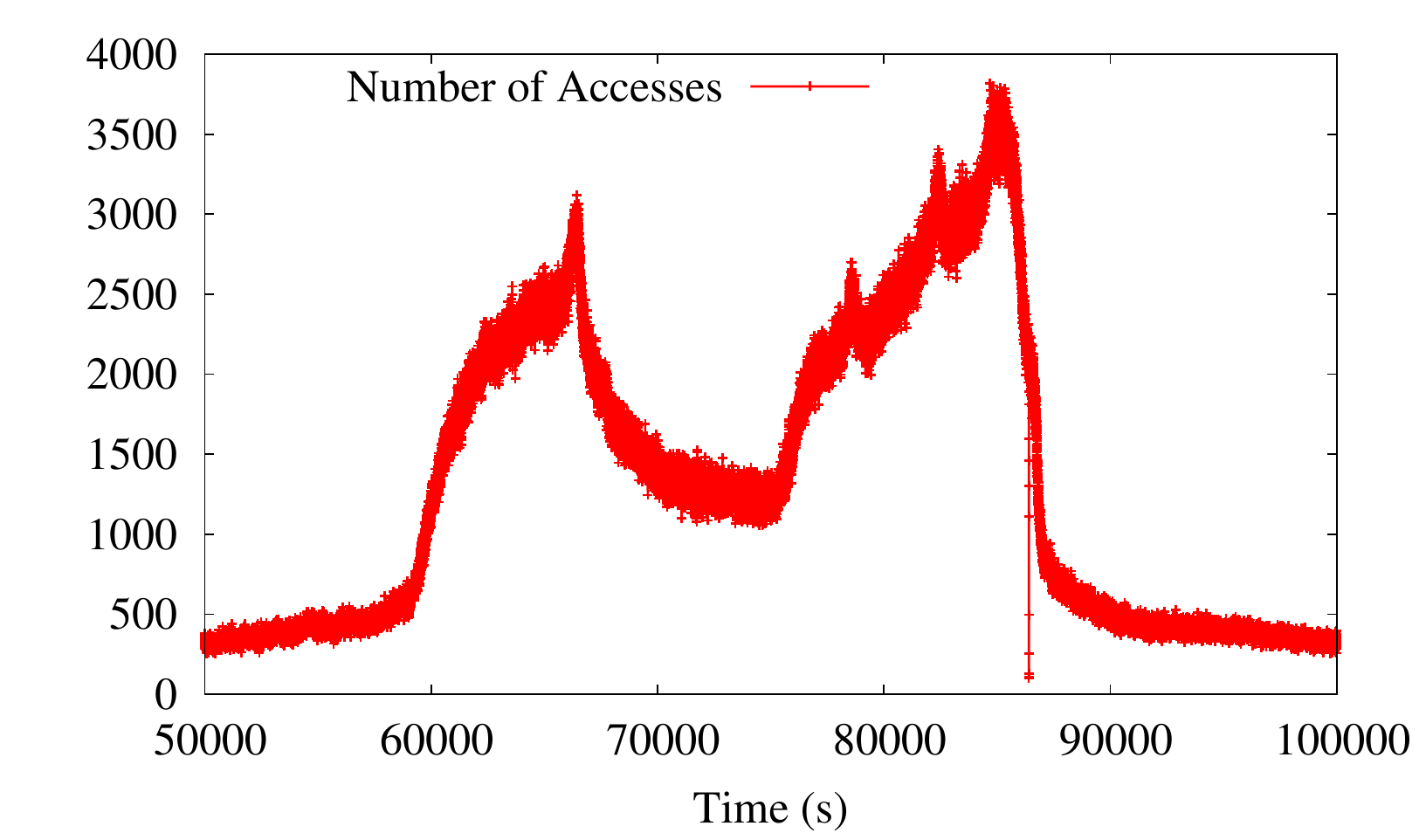} & \includegraphics[scale=0.4]{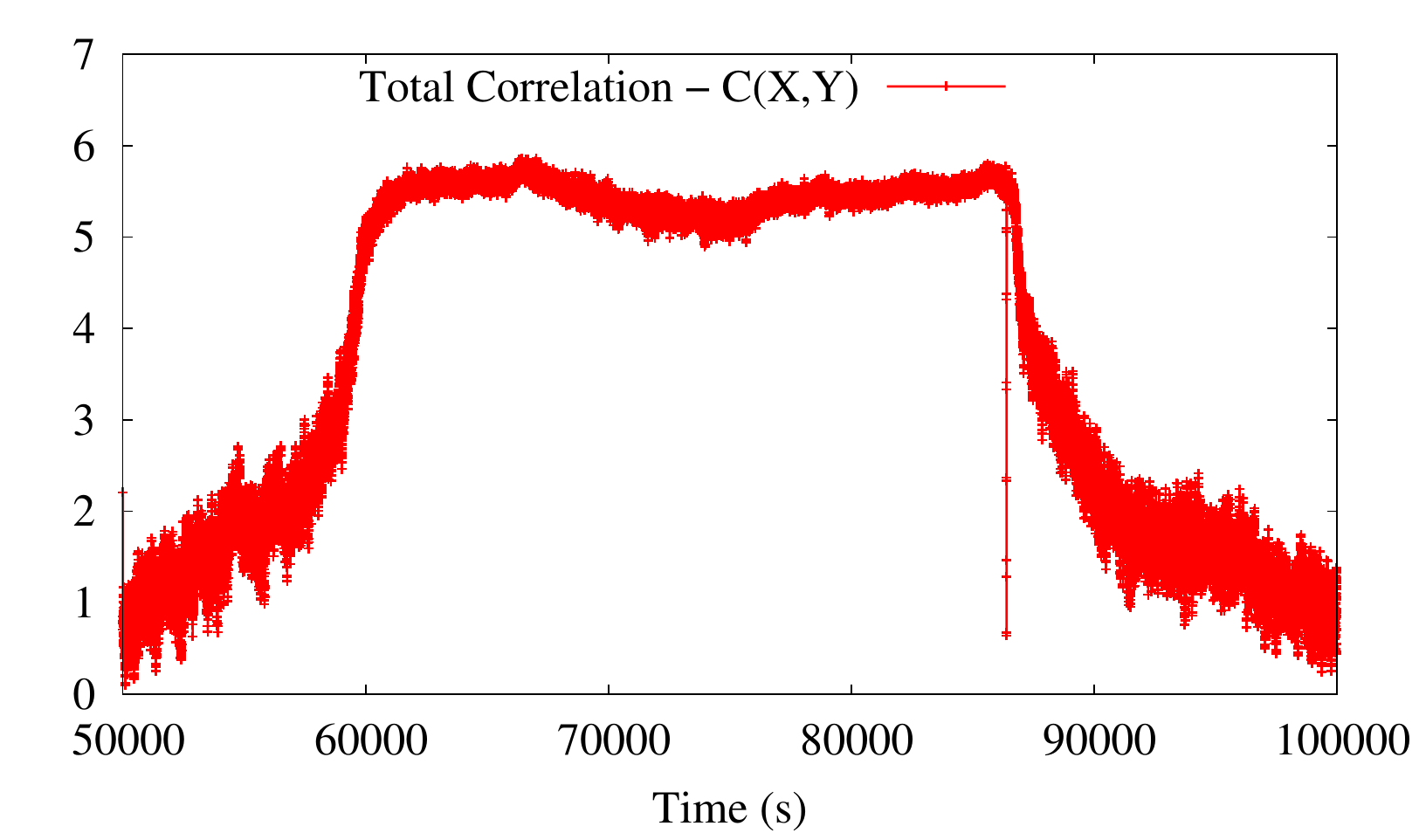} \\
(a) & (b) \\
\includegraphics[scale=0.4]{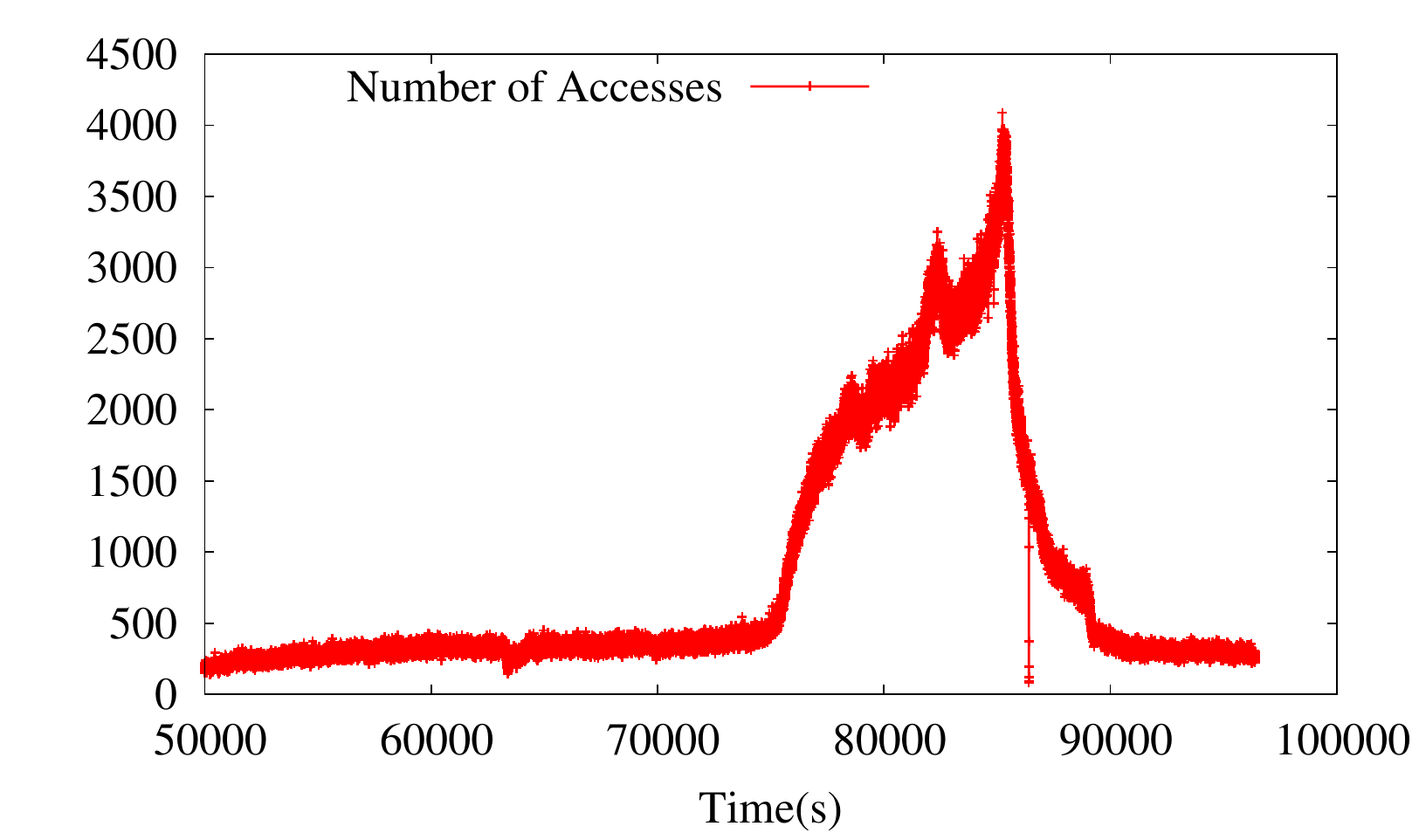} & \includegraphics[scale=0.4]{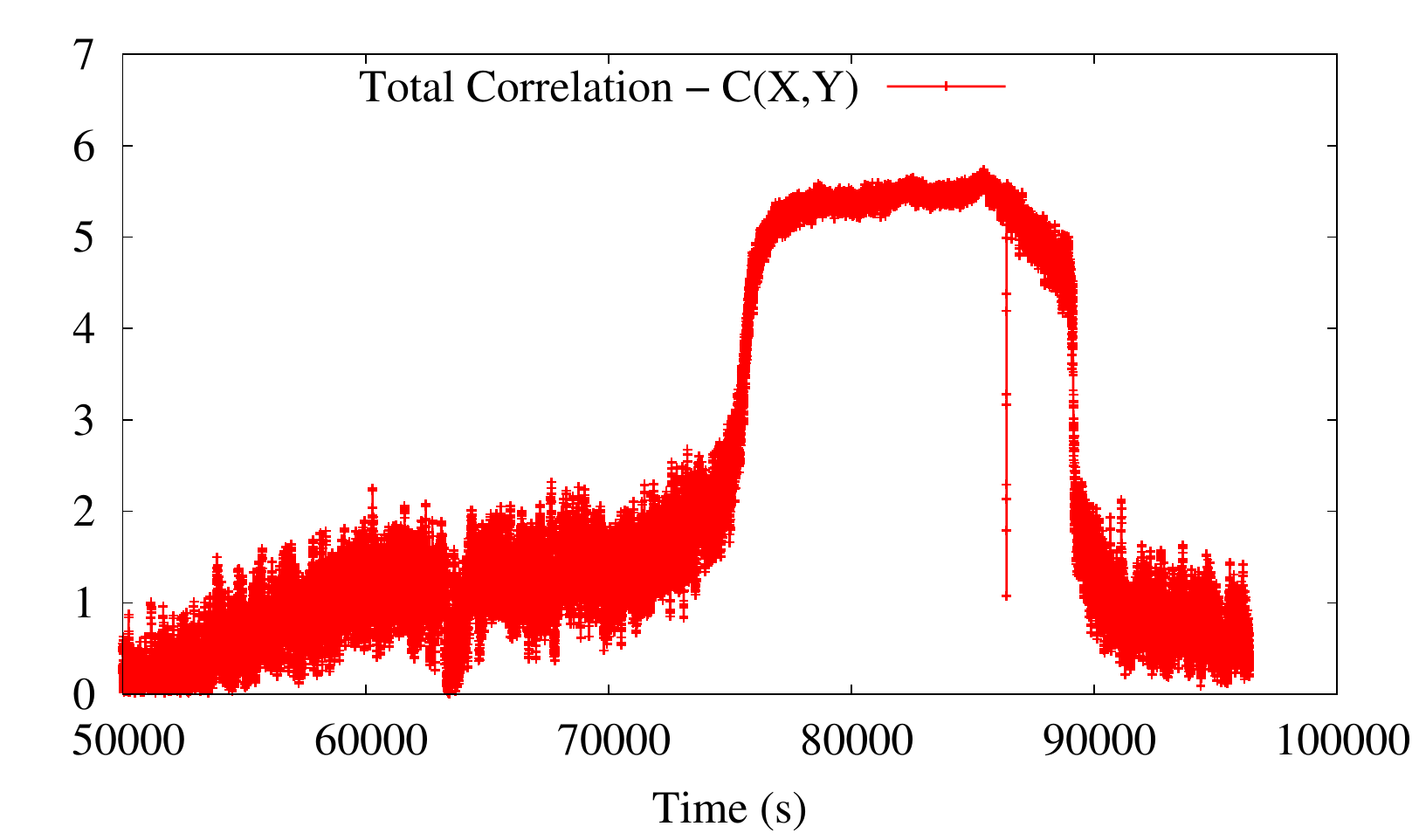} \\
(c) & (d) \\
\end{tabular}
\end{center}
\caption{Number of accesses (a),(c) and total correlation for $w=1$ (b),(d) on the 66$^{th}$ and 67$^{th}$ and 73$^{th}$ and 74$^{th}$ days of the 1998 World Cup, respectively.}
\label{fc-copa}
\end{figure}

	
	Figures \ref{fc-copa}(a) and (b) show the results for the first trace. The part (a) of this figure shows  the graphic with the number of accesses along the time.  A nontrivial flash crowd with two peaks of accesses close to the matches times can be observed. The first peak starts before 60,000s ($\sim$16:40h) and ends after 70,000s ($\sim$19:30h) and the second one starts near 75,000s ($\sim$20:50h) and approximately ends  at 90,000s ($\sim$01:00h). 
	
	Figure \ref{fc-copa}(b) presents the values of the total correlation for the first trace. The FCD mechanism correctly detected the beginning and the end of the flash-crowd event marked by substantial changes in total correlation, first up then down.
%

	As can be observed, Figure \ref{fc-copa}(c) and (d) show the same information for the second trace. The difference between these two traces is that only   one match occurred in this second one, in the 73th day. Thus,  only one peak of access can be seen in Figure \ref{fc-copa}(c), which is correctly detected by FCD in the beginning (approximately at 75,000 seconds) and in the end (at about 89,000 seconds) of the  flash-crowd event, as shown in Figure \ref{fc-copa}(d).
%

\subsection{idUFF}

	 Two idUFF traces, corresponding to two days of student enrollments in the first and second term of 2014, are used in these tests. In both  traces, one hundred different pages, each one representing one content, with one minute resolution are considered. In the first trace, ten contents suffer flash crowd, resulting in 130,992 accesses from a total of 200,661 accesses. The second one has a total of 127,599 accesses and a flash-crowd event with 67,191 accesses for only six contents. 
	 	
\begin{figure}[!ht]
\centering
\begin{center}
\begin{tabular}{c c}
\includegraphics[scale=0.4]{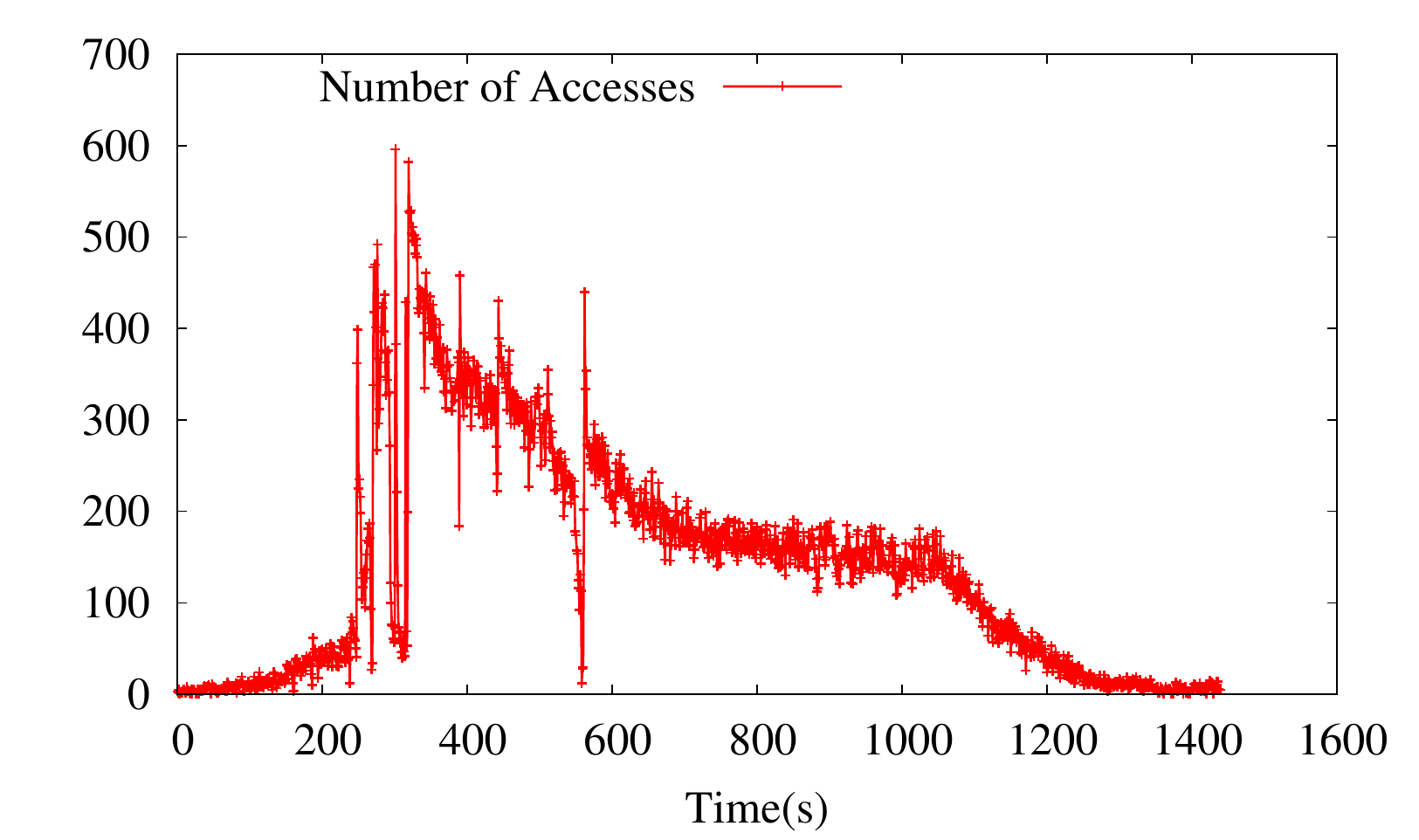} & \includegraphics[scale=0.4]{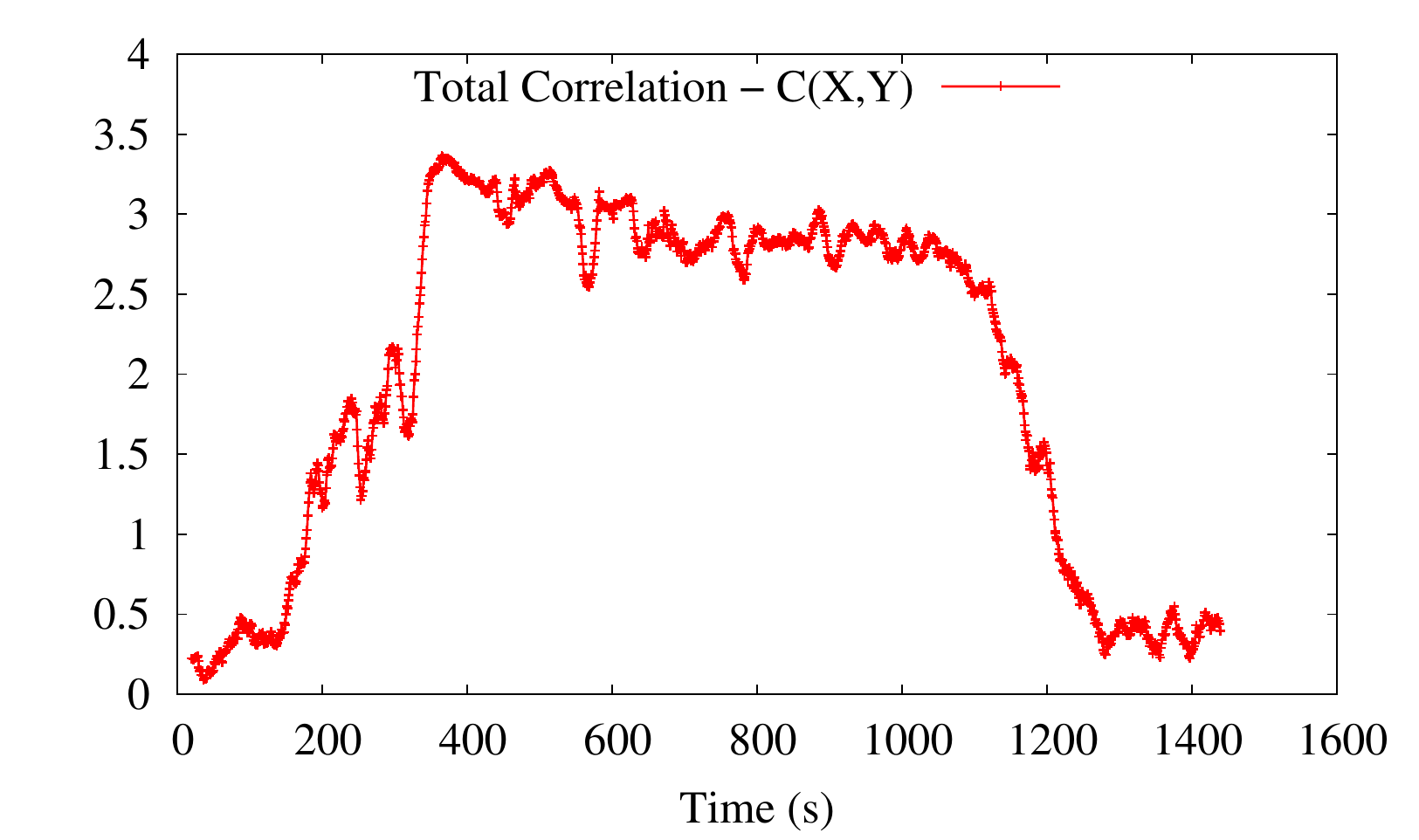} \\
(a) & (b) \\
\includegraphics[scale=0.4]{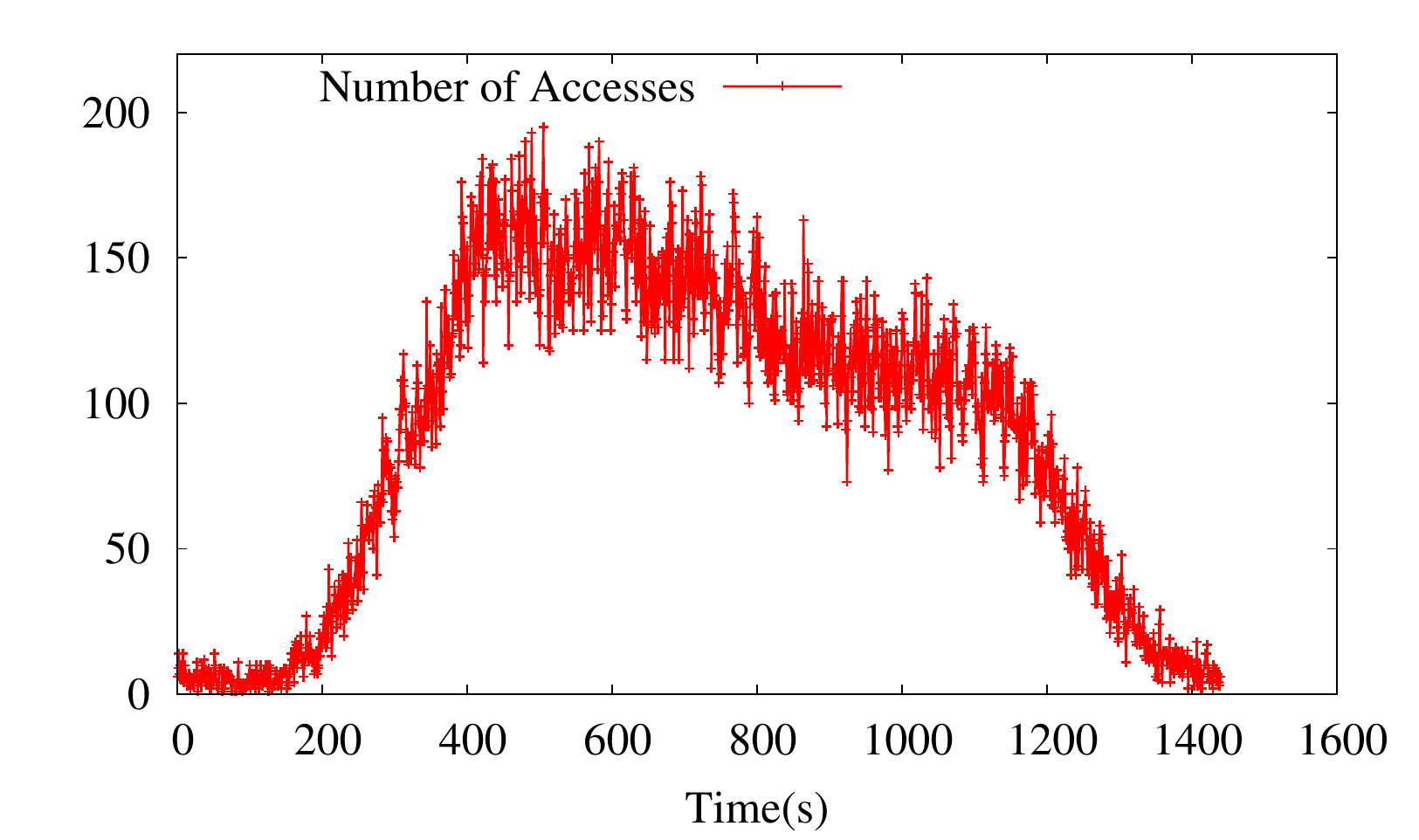} & \includegraphics[scale=0.4]{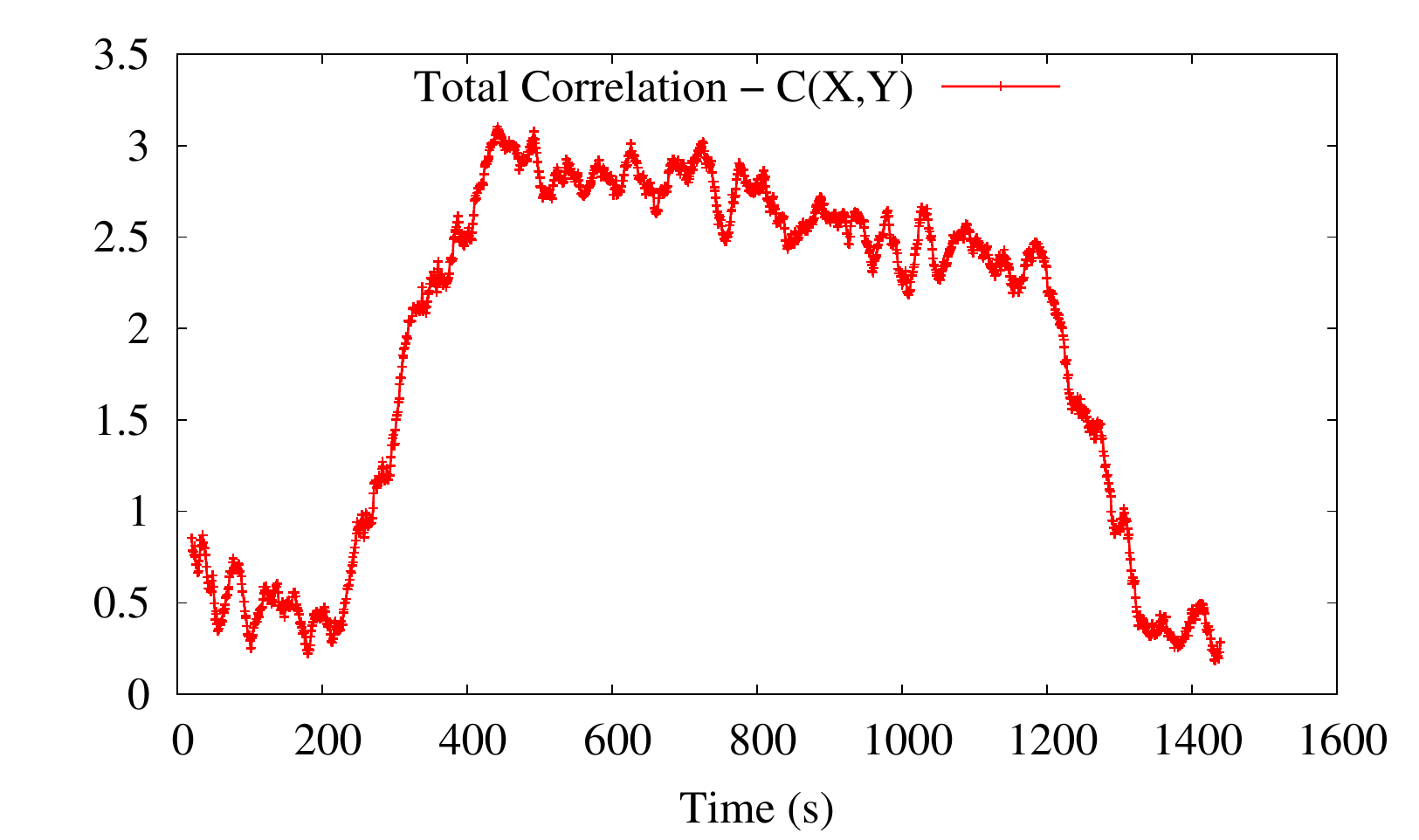} \\
(c) & (d) \\
\end{tabular}
\end{center}
\caption{Number of accesses (a),(c) and total correlation for $w=1$ (b),(d) on 07/21/2014 and 01/27/2014 in the idUFF system, respectively.}
\label{fc-iduff-all}
\end{figure}

	Figure \ref{fc-iduff-all} (a) presents the number of accesses  and  Figure \ref{fc-iduff-all} (b)  the total correlation  along   the day 01-27-14, which is  the first  one  for student enrollments. The number of accesses  increases from the 240th minute and decreases approximately at 1200 minutes. This growing was initially very unstable.  The system has halted, which affected the data collection by Google Analytics. But, independent of this variation, the FCD mechanism correctly detected the beginning and the end of the flash-crowd event as can be seen in Figure \ref{fc-iduff-all}(b).
%

Figures \ref{fc-iduff-all} (c) and (d) present the same information  concerning the day 07-21-14. Again, Figure \ref{fc-iduff-all} (b) shows that the FCD mechanism correctly detected the beginning (approximately at 200 minutes) and the end (at about 1300 minutes) of flash-crowd event.
%
%

%% file: definition.tex
\section{Handling Flash Crowds}
\label{sec:formulation}

In order to avoid that a set of contents becomes unavailable during a flash crowd it is necessary to instantiate more resources to  attend  the increasing demand for it.
There are  two possible approaches here, either to replicate  such contents in the already instantiated  Web servers or to instantiate new servers.  Replication may be not  feasible,  when there is no enough storage or bandwidth in the original  Web  servers  to attend the new demand satisfactorily. 
In this scenario, the use of cloud servers can be a more attractive approach due to the unpredictability of flash-crowd events.

\subsection{Modeling the Flash-Crowd Handling Problem}
\label{sec:model}

The Flash-Crowd Handling Problem (FCHP), introduced here, can be defined as follows. 
Let $S$ be the set of all servers (virtual or physical), where each server has a storage capacity and a maximal bandwidth.
We also consider a set of requests $R$ to be attended and a set of  contents $C$ offered by the Web application in a period of time $t\in T$.
Each request $i \in R$ requires a content and each content $k \in C$  has a fixed size. 

The FCHP is the problem of copying replicas of contents on the servers and hiring new cloud servers in order to handle the requests during the flash crowd, respecting the available storage and bandwidth and trying to minimize the  cost function defined by four sums of time: (i) the time cost $c_{i}$ to handle request $i \in R$, defined by $\sum_{i\in R}  \sum_{j \in S} \sum_{t\in T} c_{i}x_{ijt}$, where the binary variable $x_{ijt}$ indicates if the request $i$ is attended by server $j$ in period $t$, (ii) the sum of backlogging time penalty $p_{it}$ (\textit{i.e.} the penalty to postpone the attendance of request $i$ that arrived in period $t$), defined by $\sum_{i \in R} \sum_{t \in T} p_{it}b_{it}$, where $b_{it}$ indicates the postponed amount of request $i$ in period $t$, (iii) the sum of time cost $h_{k}$ to copy the content $k$, defined by $\sum_{k \in C} \sum_{j \in S} \sum_{l \in S} \sum_{t \in T} h_{k}w_{kjlt}$, where the binary variable $w_{kjlt}$ indicates if the content $k$ is copied from server $j$ to server $l$ in period $t$, and (iv) the sum of financial cost $f_j$ to hire the server $j$ in period $t$, defined by $ \sum_{j \in S} \sum_{t \in T} f_{j}z_{jt} \frac{1}{M} $, where the binary variable $z_{jt}$ indicates if the server $j$ is hired in period $t$ and $M=\max\{c_i,p_{it}b_{it},h_k,f_j\}$ $\forall i \in R, t \in T, k \in C $ and $j \in S_e$. Note that the fourth sum can not be greater than any other cost in the previous sums. So, this last sum works as a tiebreaker for solutions with the same time cost to obtain the one with the minimal financial cost.

The described scenario can be formulated as an integer programming problem, named FCHP-IP. The FCHP-IP is completed described in the Appendix.

%% file: heuristic.tex
\subsection{An ILS based heuristic for FCHP}
\label{heuristic}

One can notice that the solutions produced by the proposed mathematical formulation can not be used in practice. The reason for that lies in the fact that the formulation uses future knowledge about content requests. Moreover, exact procedures have often proved incapable of finding solutions as they are extremely time-consuming, particularly for real-world problems. 
%

In this context, we have designed and implemented an ILS-RVND heuristic \cite{ils}. The ILS-RVND can be defined as a multi-start method that uses (i) a random/greed heuristic in a constructive phase, (ii) a Variable Neighborhood Descent with Random neighborhood ordering (RVND) in the local search phase and (iii) perturbations moves as  a diversification mechanism. 
The main steps of the ILS-RVND are described in Algorithm \ref{alg_ils-rvnd}. The multi-start method executes $iter\_max$ iterations, where at each iteration the constructive procedure generates an initial feasible solution $s$ (line 2) that may be improved
by the RVND method (line 3). The internal loop (lines 5-14) aims to improve the initial solution by "shaking" solution $s$ with a perturbation mechanism (line 7) and re-applying the local search method. The parameter $level\_max$ represents the maximum level of perturbation applied to the current solution. Next, we provide a short explanation of the $3$ main components of the ILS-RVND heuristic applied to FCHP.

\begin{algorithm}[thp]
\caption{\footnotesize ILS-RVND  \label{alg_ils-rvnd}}
	\For{$i:=1$ \textbf{to} $iter\_max$}{
		$s:=constructive\_phase()$;\\
		$s:=RVND(s)$;\\
		$level:=0$;\\
			\While{$level<level\_max$}{
			$s':=s$;\\	
			$s':=perturbation(s', level)$;\\
			$s':=RVND(s')$;\\
			\eIf{$f(s')<f(s)$}{
				$s:=s'$; $level:=0$;\\
			}
			{
				$level:=level+1$;\\
			}
	}
	\If{$f(s)<f(s^*)$}{
		$s^*:=s$;\\
	}
	}
\end{algorithm}


The constructive phase consists of a greedy heuristic that builds a feasible solution by iteratively assigning each request to the cheapest server with enough bandwidth. The chosen server must hold the required content or have enough storage to keep it. The method first tries to exhaust the set of Web application servers before starts hiring servers available in the cloud. 
When there is no available server capable of attending the request, including the cloud servers that could be hired, one of the following approaches is applied. If there is a server with enough bandwidth and no available storage, a content  is removed from it according to the {\em Least Recently Used} $(LRU)$ strategy. Otherwise, the request attendance is postponed and a  backlog cost is added to the  associated cost function. The randomization of this heuristic is achieved by creating a random order of the  requests.


The local search is executed by a VND algorithm \cite{Mladenovic1997} with a random neighbourhood ordering (RVND). 
Let $N$ be an unordered set of neighbourhood structures. Whenever one neighbourhood fails to improve the current solution, the RVND randomly chooses another neighbourhood in $N$ to continue the search. The local search halts when no better solution is found in the set of neighbourhood structures of the current solution.
%
%
In order to describe the neighbourhood structures, we need some additional notation. We define
a solution $\{(k_1,j_1,\mathcal{R}_1,t_1),$ $(k_2,j_2,\mathcal{R}_2,t_2),...\}$ as a set of $4-$tuples $(k,j,\mathcal{R},t)$ representing
 that content $k$ is replicated in server $j$ to attend a set of requests $\mathcal{R}$ on period $t$. The ILS-RVND is composed by the following five neighbourhoods:
%
\begin{itemize}
	\item {\bf Shift} $(k,j_a,\mathcal{R},t) \rightarrow (k,j_b,\mathcal{R},t)$: transfer one tuple from a server $j_a$ to a server $j_b$. 
	\item {\bf Swap} $(k_a,j_a,\mathcal{R}_a,t_a)$, $(k_b,j_b,\mathcal{R}_b,t_b)$ $\rightarrow (k_a,j_b,\mathcal{R}_a,t_a)$, \\$ (k_b,j_a,\mathcal{R}_b,t_b)$: one tuple from a server $j_a$ is permuted with a tuple from server $j_b$.
	\item {\bf Split} $(k,j_a,\mathcal{R}_a,t)$ $\rightarrow$ $(k,j_b,\mathcal{R}_b,t)$, $(k,j_c,\mathcal{R}_c,t)$: split requests from server $j_a$ between servers $j_b$ and $j_c$, where $\mathcal{R}_a=\mathcal{R}_b \cup \mathcal{R}_c$. 
	\item {\bf Merge} $(k,j_a,\mathcal{R}_a,t), (k,j_b,\mathcal{R}_b,t) \rightarrow (k,j_c,\mathcal{R}_a \cup \mathcal{R}_b,t)$: merge set of requests from servers $j_a$ and $j_b$ into a new tuple in server $j_c$.
	\item {\bf $d-$Delay} $(k,j,\mathcal{R},t) \rightarrow (k,j,\mathcal{R},t+d)$: delay a tuple in $d$ periods of time. Note that we can have a positive delay $(d > 0)$ or a negative delay $(d < 0)$. 
\end{itemize}
It is important to emphasize that only feasible movements are accomplished. In addition, as the Swap neighbourhood is the most expensive one, we analyzed only $5\%$ of possible movements.
%
%
With respect to the perturbation mechanism, we perform multiple Shift, Swap, Split and Merge movements randomly chosen in such a way that the resulting modification is sufficient to escape from local optima and analyse different regions of the search space.

%% file: results_FCHP.tex
\section{Experimental Results for Handling Flash Crowds}
\label{sec:result2}

In this section, we present results for the comparison between the FCHP-IP mathematical formulation and the FCHP-ILS heuristic. 
	 We evaluate FCHP-ILS, in terms of  quality of solution and execution time,  by comparing it with the solutions given by the formulation FCHP-IP when solved with the CPLEX $12.5.1$ \cite{cplex}. Tests were run  in a computer with a processor Intel Core i7-3820 3.60GHz  and  a 32Gb memory running Ubuntu $14.04$. The  FCHP-ILS was implemented in the Programming language C/C++, gcc version $4.8.2$.
	
	 The programs were run over a set of instances created from a  real trace obtained from the 1998 World Cup site \cite{copa98}.  Because this trace presents a huge number of requests, the total time of the trace was discretized in hours and only the requests for the ten most accessed contents in each interval of time were considered. Thus, instances with reduced size, but still presenting flash-crowd events, were created. Remark that the original instance could not be solved in a reasonable time by the CPLEX. 

	 Table \ref{Instance} presents for each of the twelve created instances, the  associated number of contents, requests and number of periods (hours). The last two instances were created by the synthetic trace generator and used for the real experiments, described in the next section.
	
\begin{table}[htp]
\small
\setlength{\tabcolsep}{3mm}
\begin{center}
\tbl{Instance description.\label{Instance}}{
\begin{tabular}{|c|r|r|r|}
\hline 
Instance & \# of Contents  & \# of Requests & Periods (hours) \\
\hline
1	&	44	&	1798		&	12		\\
2	&	60	&	1696		&	12	    \\
3	&	88	&	983	    	&	12		\\
4	&	54	&	3546		&	24		\\
5	&	98	&	3427		&	24		\\
6	&	161	&	1922		&	24		\\
7   &   47  &   4327	    &   36      \\
8   &   88  &   4073        &   36      \\
9   &   104  &   1676        &   36      \\
10   &  69  &   7110 		&   48      \\
11   &   3  &   105         &   60      \\
12   &   4  &   186         &   60      \\
\hline
\end{tabular}}
\end{center}
\end{table}
	
Table \ref{teorico} shows the results obtained by FCHP-IP and FCHP-ILS. The first column identifies the instance. The following five columns present the results obtained by FCHP-IP: number of hired  on demand servers, the total cost, the attendance cost, replication cost and the execution time to obtain the optimum solution. Following, the next columns present the same results of FCHP-ILS. Finally, the last column, shows the gap between the solutions given by FCHP-ILS and FCHP-IP. The  values shown for FCHP-ILS are averages of ten executions, where  $2$, $1$ and $1$ were used for the number of iterations, the number of perturbations and the value of $d$ ({\bf $d-$Delay} neighbourhood), respectively. 

\begin{table}[htp]
\small
\begin{center}
\setlength{\tabcolsep}{1.1mm}
\tbl{Results of FCHP-ILS Metaheuristic and FCHP-IP Mathematical Formulation using CPLEX.\label{teorico}}{
\begin{tabular}{|c|crrrr|crrrr|c|}
\hline 
\multirow{3}{1cm}{Instance} & \multicolumn{5}{|c|}{FCHP-IP} & \multicolumn{5}{|c|}{FCHP-ILS} & \multirow{3}{1cm}{Gap(\%)}\\
\cline{2-11}
 & Serv & \multicolumn{3}{|c|}{Time Cost} & & Serv & \multicolumn{3}{|c|}{Time Cost} & & \\
 &  OD & \multicolumn{1}{|c}{Total} & Attend & Repli & \multicolumn{1}{|c|}{Time (s)} & OD & \multicolumn{1}{|c}{Total} & Attend & Repli  & \multicolumn{1}{|c|}{Time (s)} &\\
\hline 
1 & 19 & 22600.5 & 18979.6 & 3620.9 & 99.7 & 5 & 22872.9 & 19023.6 & 3849.3 & 2.0 & 1.2 \\
2 & 3 & 22990.6 & 18064.2 & 4926.4 & 92.1 & 4 & 23305.7 & 18066.2 & 5239.5 & 2.2 & 1.4 \\
3 & 14 & 17347.6 & 10151.6 & 7196.0 & 32.4 & 2 & 17796.4 & 10151.6 & 7644.8 & 1.6 & 2.6 \\
4 & 49 & 45846.3 & 38084.2 & 7762.1 & 506.8 & 9 & 46101.3 & 38063.8 & 8037.5 & 13.8 & 0.6 \\
5 & 30 & 49520.3 & 35495.4 & 14024.9 & 670.1 & 5 & 49989.5 & 35464.8 & 14524.7 & 13.6 & 0.9 \\
6 & 0 & 42595.2 & 19604.4 & 22990.8 & 91.1 & 0 & 43416.3 & 19604.4 & 23811.9 & 11.0 & 1.9 \\
7 & 17 & 55749.6 & 46095.4 & 9654.2 & 2518.4 & 12 & 56020.8 & 46122.4 & 9898.4 & 23.3 & 0.5 \\
8 & 11 & 60017.3 & 42019.6 & 17997.7 & 963.7 & 7 & 60489.1 & 42041.1 & 18448.0 & 20.3 & 0.8 \\
9 & 6 & 38356.3 & 17130.2 & 21226.1 & 134.8 & 4 & 38892.0 & 17115.2 & 21776.8 & 8.0 & 1.4 \\
10* & 57 & 93812.5 & 75412.0 & 18400.5 & 2723.0 & 15 & 94144.2 & 75412.0 & 18732.2 & 85.9 & 0.4 \\
11* & 440 & 28451.7 & 21957.4 & 6494.3 & 7388.8 & 115 & 29577.9 & 21957.4 & 7620.5 & 16.2 & 4.0 \\
12* & 239 & 41228.2 & 32208.9 & 9019.3 & 14382.2 & 214 & 43599.7 & 32208.9 & 11390.8 & 24.1 & 5.8 \\
\hline
Average &  &  &  &  & 2466.9 &  &  &  &  & 18.5 & 1.8 \\
\hline 
\end{tabular}}
\end{center}
\end{table}
	
	In this table, we can observe that for all instances the FCHP-ILS obtains good solutions with small gaps, in average $1.8\%$. Moreover, it takes significantly less time when compared with the FCHP-IP when solved with the CPLEX, in average $18.5$s against $2466.9$s, respectively. 
	The FCHP problem extends the Replica Placement Problem (RPP), which belongs to the NP-hard class \cite{aioffi05}. Thus, as the input data increases, the FCHP-IP is not capable of proving the optimality of the solution in a reasonable time. In addition, FCHP-IP needs a higher memory capacity to solve bigger instances. The results for the last three instances confirm this, where the CPLEX could not prove the optimality of the solution due to lack of memory. Remark that the two last instances were used in the real experiments, presented next.

%% file: results_FCD_FCHP.tex
\section{Experimental Results for Detecting and Handling Flash Crowds}
\label{sec:result3}

	In this section, we present the results for six tests executed in real small-scale scenarios, using the Amazon Cloud. The first two tests are the comparison between the FCHP-ILS heuristic and the combination of Amazon's Auto Scaling (AS) and Load Balancing mechanisms (LB) \cite{AS,ELB}. In the next four tests, we introduce the FCD mechanism to detect the flash-crowd events. Thus, these tests are the comparison between the combination of FCHP-ILS heuristic and FCD mechanism and the combination of Amazon’s AS and LB mechanisms.
	
	In Table \ref{scenarios}, the descriptions for all scenarios are presented. The first column identifies the scenario. The next four columns present the number of contents and the intervals of the \textit{ramp-up}, \textit{sustained traffic} and  \textit{ramp-down} phases (in seconds). Note that for the second, fourth and sixth scenarios, the table has two rows for the flash-crowd phases indicating the occurrence of two flash crowd events.
		
\begin{figure}[ht]
\centering
\begin{center}
\begin{tabular}{c c}
\includegraphics[scale=0.4]{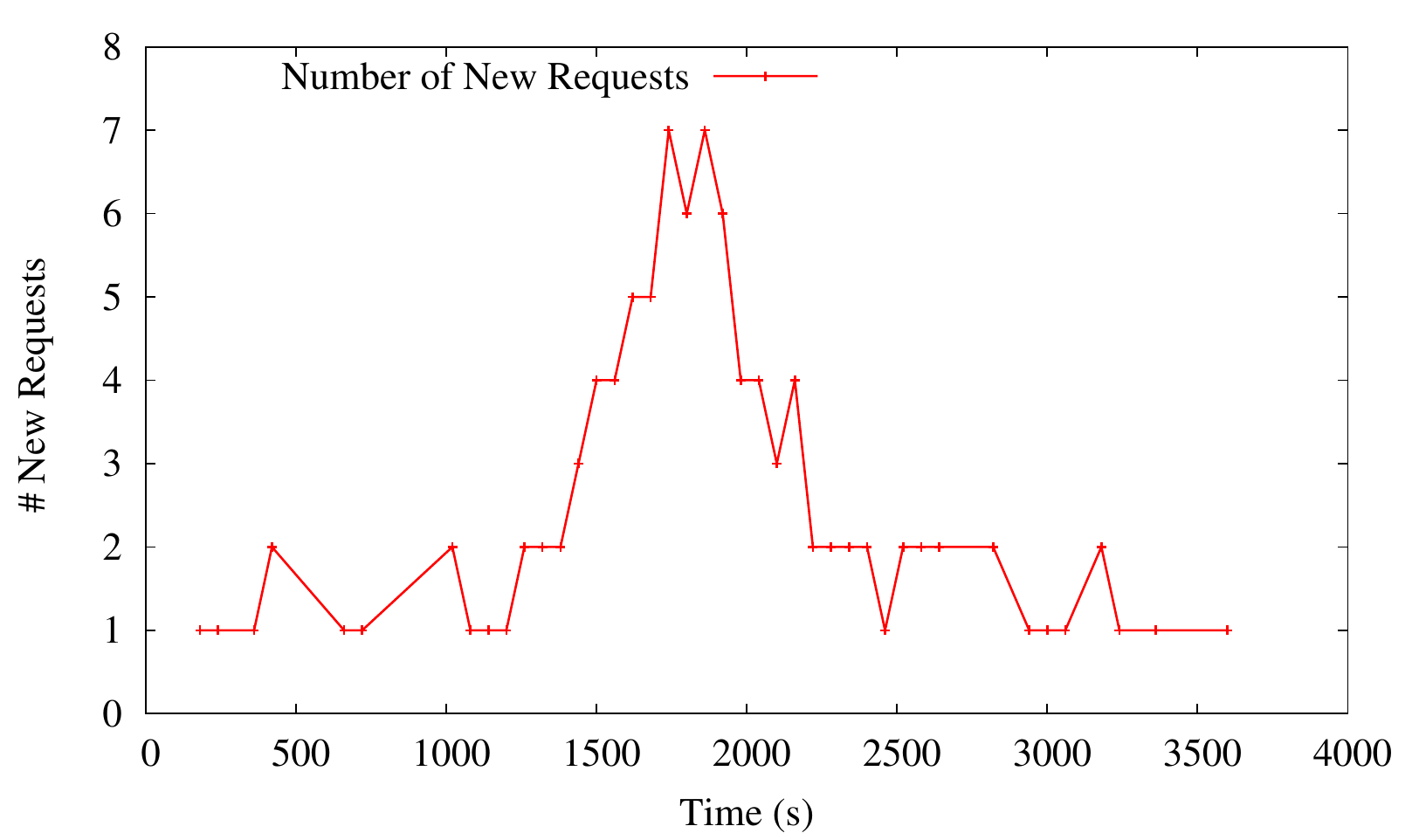} & \includegraphics[scale=0.4]{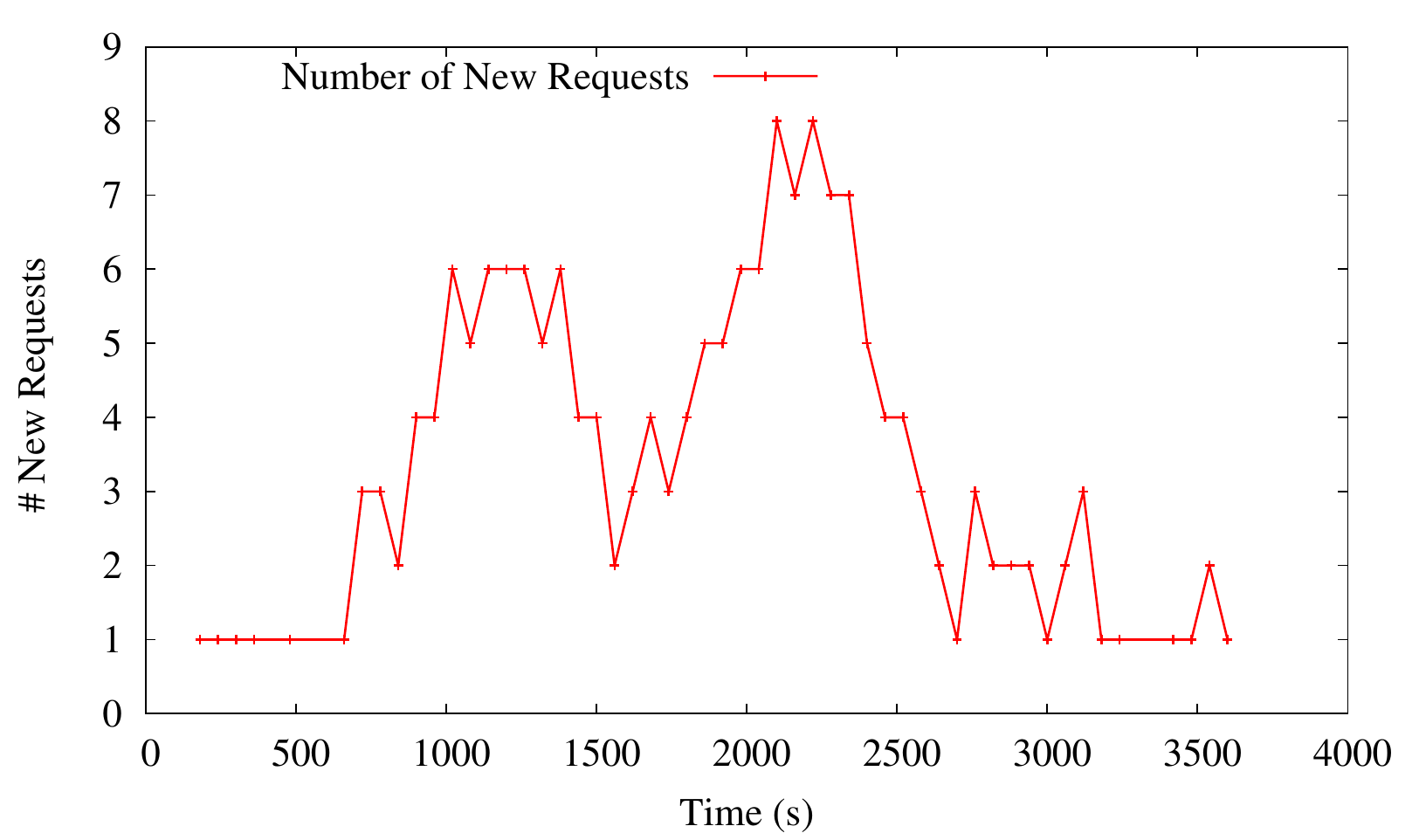} \\
(a) & (b) 
\end{tabular}
\end{center}
\caption{(a) New generated requests per second for the (a) first and (b) second scenario.}
\label{fc-figure}
\end{figure}

	In the first scenario, the Web application offers  three different contents  with  the following sizes: $1.9$, $1.5$ and $0.9$ Gbytes, but only one content was involved in the flash crowd. Figure \ref{fc-figure}(a) presents the number of requests along the time. 
	%
	In the second scenario, the Web application offers four different contents  with  the following sizes: $0.8$, $1.0$, $1.5$ and $2.0$ Gbytes. In this scenario there are two flash-crowd events. The requests for this scenario are shown in Figure \ref{fc-figure}(b).
		
	In these two first tests, the servers' bandwidth is  $10$ Mbytes/s while  the clients' bandwidth  is $1$ Mbytes/s  bandwidth  in average. The content of $0.9$ Gbytes was the only one involved in the flash crowd for the first test and the contents of $0.8$ and $1.0$ Gbytes for the second one. In addition, the Web application started by using two virtual machines of type \textit{m3.large}  of the Amazon, each one containing all the contents.
		
\begin{figure}[ht]
\centering
\includegraphics[scale=0.3]{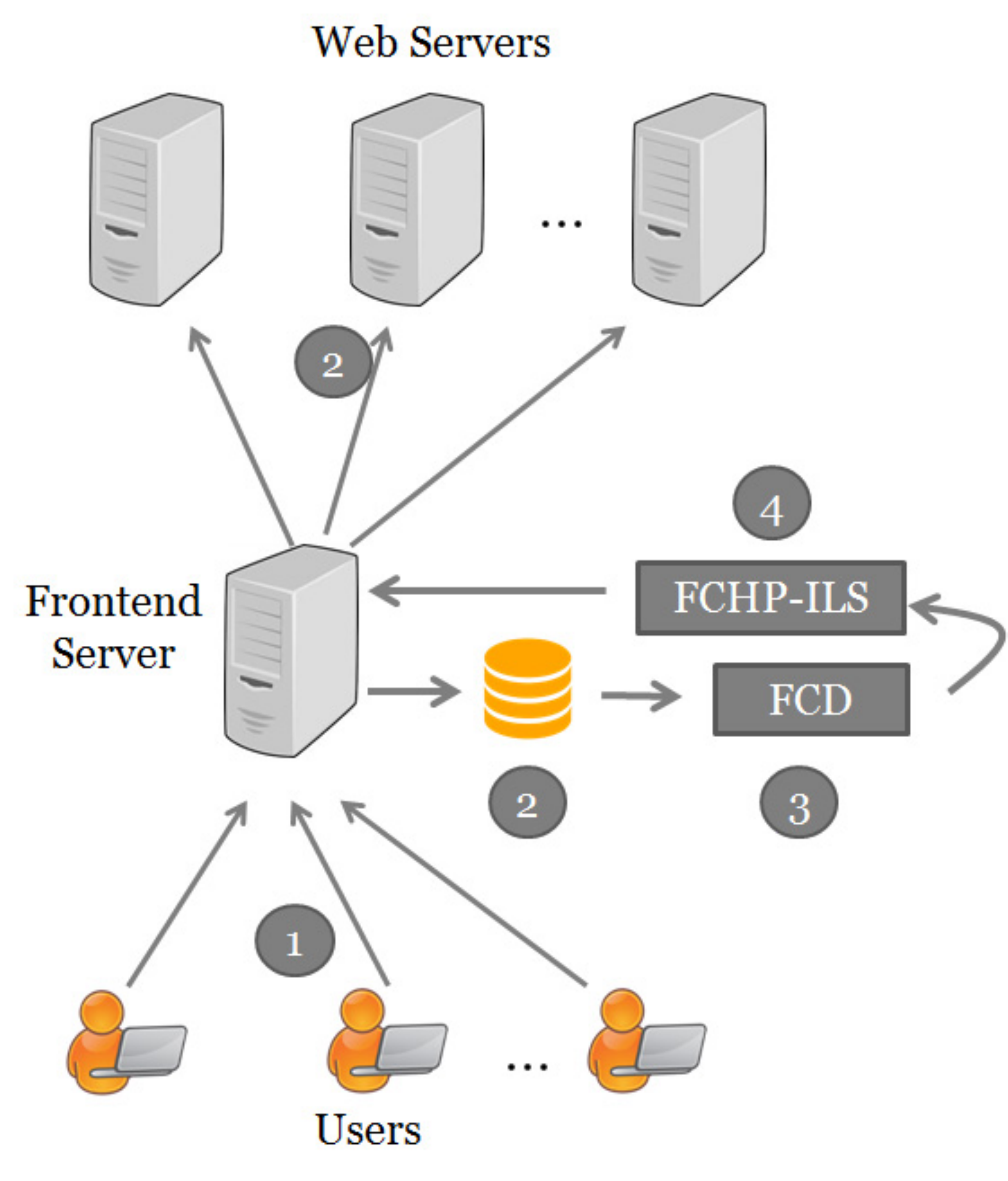}
\caption{Combination of FCD and FCHP-ILS mechanisms.}
\label{detec+ils}
\end{figure}

	Before describing the next scenarios, we need to explain the the combination of the proposed solutions for the detection and mitigation of flash-crowd events, which is illustrated in Figure \ref{detec+ils}. At first, in step (1), users request contents from the Web application. In step (2), the frontend server redirects the requests to the appropriate server and registers these requests in the log base. This frontend server is similar to the server that runs the Auto Scaling and Load  Balancing mechanisms, in Amazon. Next, in step (3), the FDC mechanism is executed periodically over the log base. Finally, in step (4), if a Flash Crowd is detected, the FCHP-ILS heuristic is executed. Note that until the detection of the end of the flash-crowd event, the FCHP-ILS is executed periodically to update current solution.
	
 In these scenarios, the clients' bandwidth is $250$ Kbytes/s in average. For the third and the sixth, the servers' bandwidth is $8$ Mbytes/s while for the fourth and the fifth is $10$ Mbytes/s. Due to similarity, the figures showing the number of accesses and the values of the total correlation along the time  are presented only for two scenarios, with one and two flash-crowd events and different content sizes.
	
\begin{figure}[ht]
\centering
\begin{center}
\begin{tabular}{c c}
\includegraphics[scale=0.4]{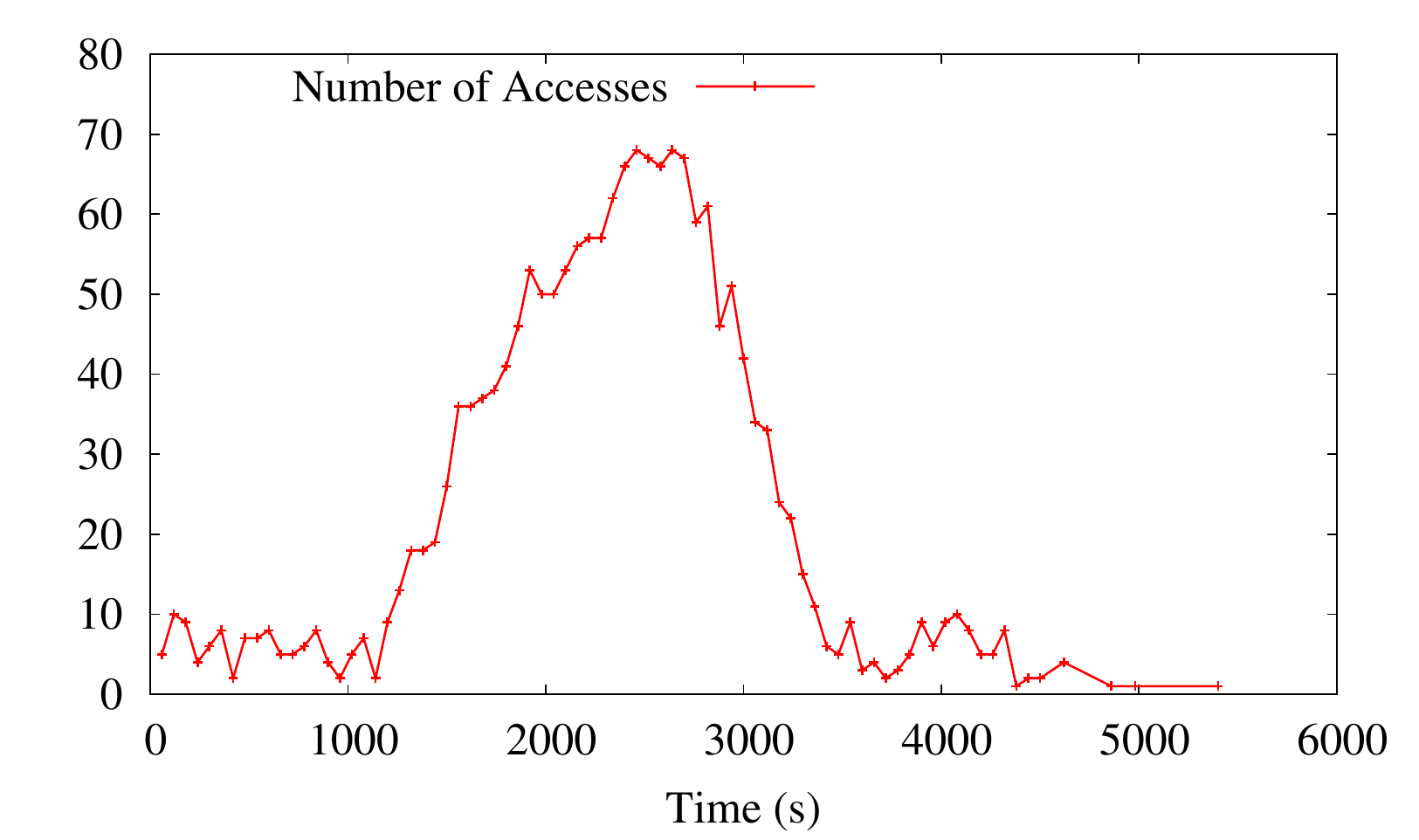} & \includegraphics[scale=0.4]{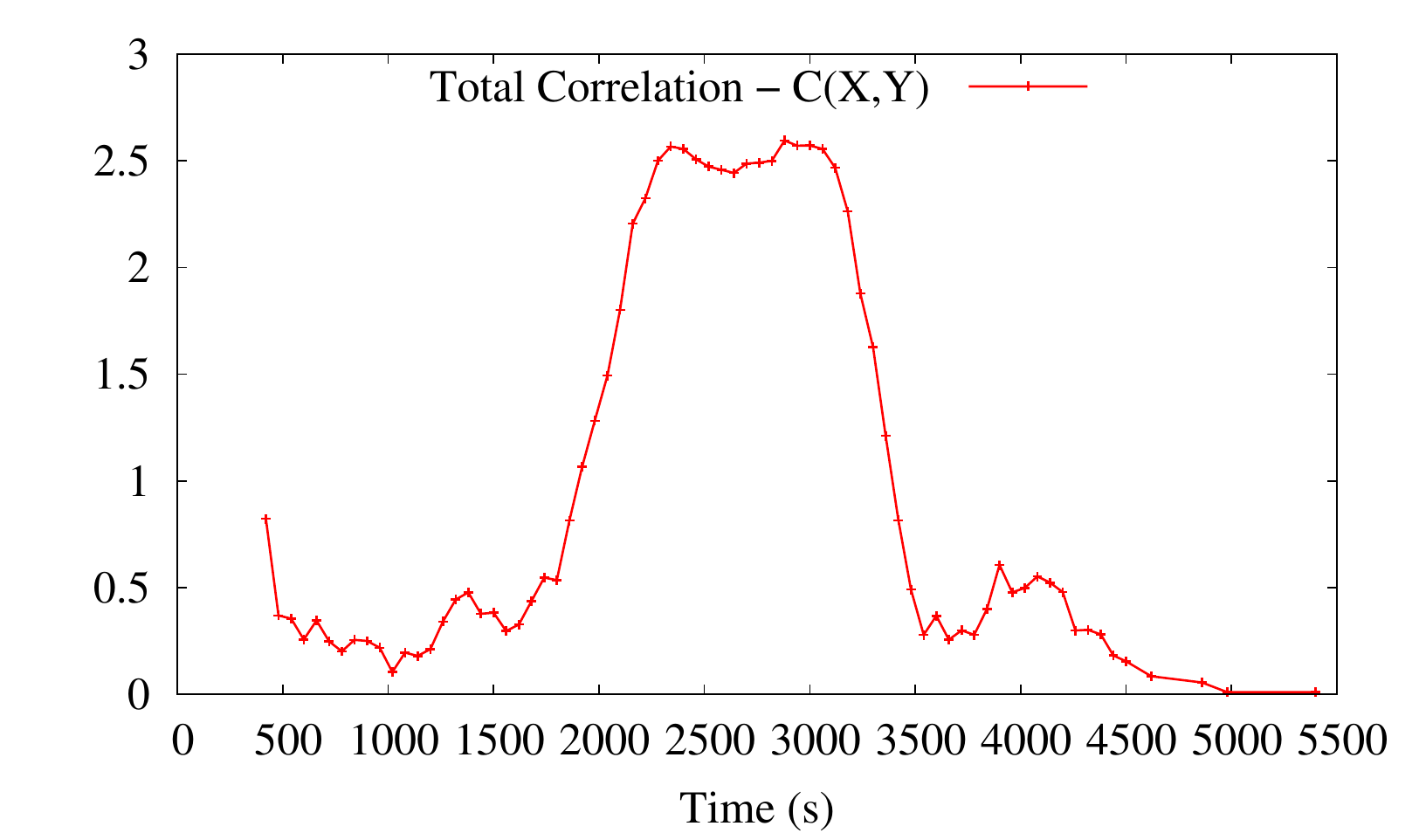} \\
(a) & (b) \\
\includegraphics[scale=0.4]{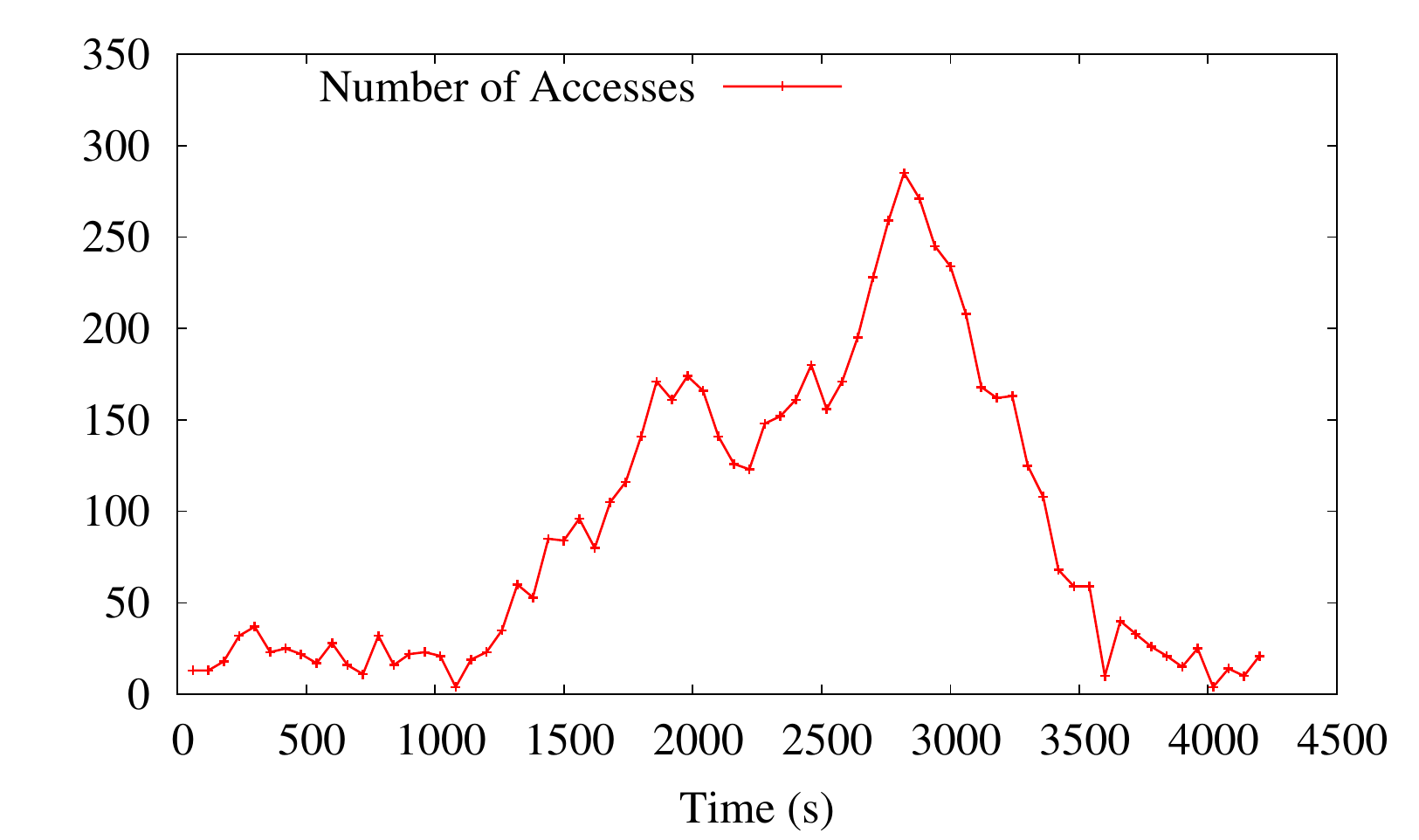} & \includegraphics[scale=0.4]{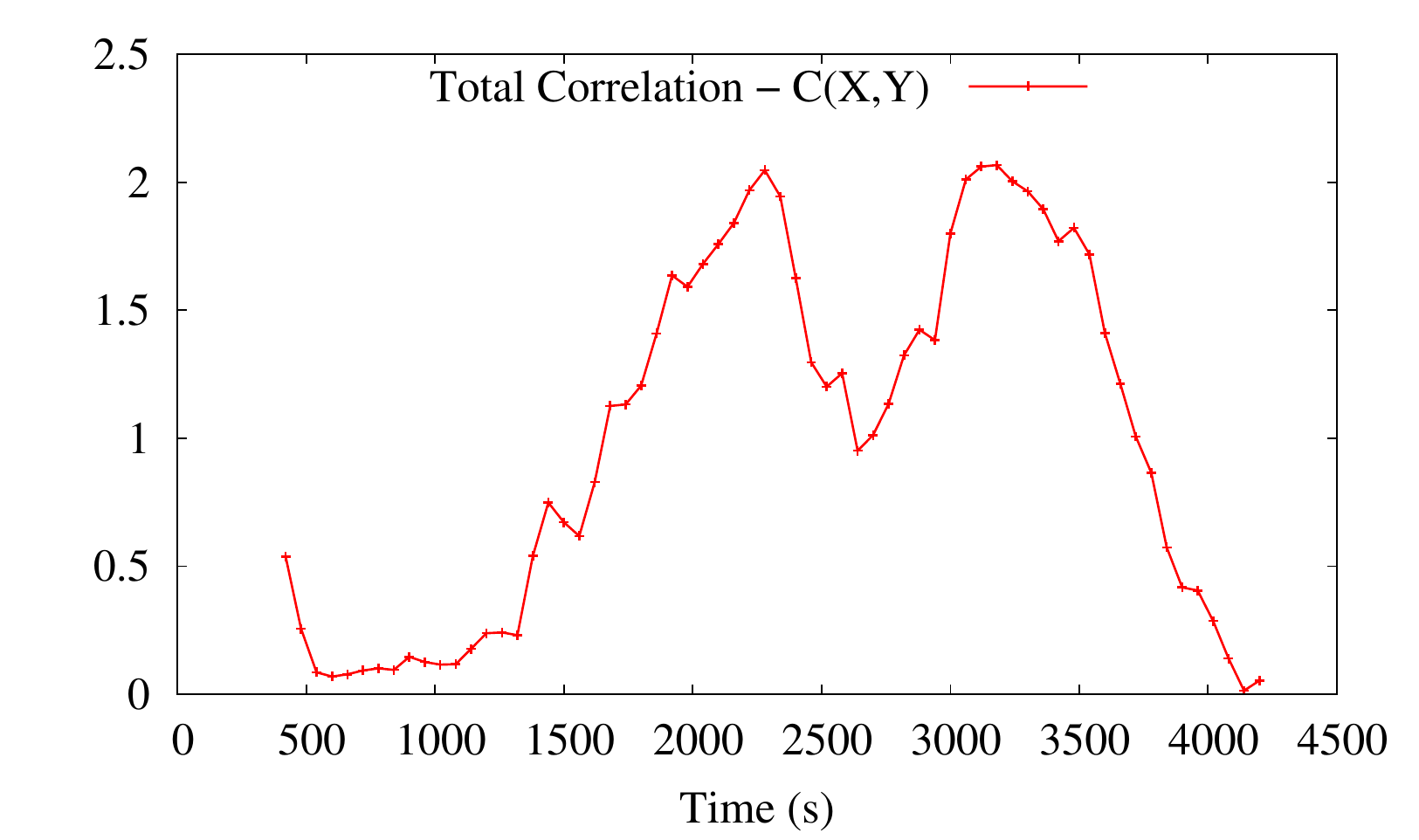} \\
(c) & (d) \\
\end{tabular}
\end{center}
\caption{Number of accesses (a),(c) and total correlation for $w=1$ (b),(d) for the third and sixth scenario, respectively.}
\label{fc-fcd-ils-all}
\end{figure}

	The third scenario contains contents of sizes $2,4,8,16,32,64,128,256,512$ and $1024$ Mbytes, with two different contents for each size, totalling twenty different contents. Figures \ref{fc-fcd-ils-all}(a) and Figure \ref{fc-fcd-ils-all}(b) present the number of requests and values of the total correlation for this scenario along the time, respectively.
%
	
	The fourth scenario also has contents of sizes $2,4,8,16,32,64,128,256,512$ and $1024$ Mbytes, with two different contents for each size, totalling twenty different contents. In this scenario there are two flash-crowd events. The fifth one contains twenty different contents with average sizes of $10$ Mbytes, but only one flash crowd.
	Finally, the sixth scenario also contains twenty different contents with average sizes of $10$ Mbytes. In this scenario there are two flash-crowd events. The accesses for this scenario are shown in Figure \ref{fc-fcd-ils-all}(c) and the values of the total correlation along the time, in Figure \ref{fc-fcd-ils-all}(d). 
	
	For all scenarios, the FCD mechanism detected the flash-crowd events marked by an increase and a decrease in the total correlation during the \textit{ramp-up phase} and the \textit{ramp-down phase}, respectively. Remark that in the scenarios with two flash crowds, the FCD detected correctly its beginning in the \textit{ramp-up phase} of the first event and its end in the \textit{ramp-down phase} of the last one.

\begin{table}[htp]
\small
\begin{center}
\tbl{Scenarios description.\label{scenarios}}{
\begin{tabular}{|c|c|c|c|c|}
\hline 
\multicolumn{1}{|c|}{\multirow{2}{*}{Scenario}} & \multicolumn{1}{|c|}{\multirow{2}{*}{\# of Contents}} & \textit{Ramp-up}  & \textit{Sustained traffic} & \textit{Ramp-down} \\
& & Phase (s) & Phase (s) & Phase (s)\\
\hline
\multicolumn{1}{|c|}{\multirow{2}{*}{1}} & \multicolumn{1}{|c|}{\multirow{2}{*}{3}} & \multicolumn{1}{|c|}{\multirow{2}{*}{[$1140$,$1740$]}} & \multicolumn{1}{|c|}{\multirow{2}{*}{($1740$,$1860$)}} & \multicolumn{1}{|c|}{\multirow{2}{*}{[$1860$,$2460$]}}\\
&&&&\\
\hline
\multicolumn{1}{|c|}{\multirow{2}{*}{2}} & \multicolumn{1}{|c|}{\multirow{2}{*}{4}}  & [$540$,$1140$] & ($1140$,$1320$) & [$1320$,$1860$]\\
 &  & [$1500$,$2100$] & ($2100$,$2220$) & [$2220$,$2820$]\\
 \hline
\multicolumn{1}{|c|}{\multirow{2}{*}{3}} & \multicolumn{1}{|c|}{\multirow{2}{*}{20}}  & \multicolumn{1}{|c|}{\multirow{2}{*}{[$1200$,$2340$]}} & \multicolumn{1}{|c|}{\multirow{2}{*}{($2340$,$2700$)}} & \multicolumn{1}{|c|}{\multirow{2}{*}{[$2700$,$3300$]}} \\
&&&&\\ 
\hline
\multicolumn{1}{|c|}{\multirow{2}{*}{4}} & \multicolumn{1}{|c|}{\multirow{2}{*}{20}} & [$1140$,$1860$] & ($1860$,$1980$) & [$1980$,$2700$]\\
&&  [$2160$,$2820$] &  ($2820$,$2860$) & [$2860$,$3600$]\\
\hline
\multicolumn{1}{|c|}{\multirow{2}{*}{5}} & \multicolumn{1}{|c|}{\multirow{2}{*}{20}} & \multicolumn{1}{|c|}{\multirow{2}{*}{[$1140$,$1800$]}} & \multicolumn{1}{|c|}{\multirow{2}{*}{($1800$,$1860$)}} & \multicolumn{1}{|c|}{\multirow{2}{*}{[$1860$,$2460$]}} \\
&&&&\\
\hline
\multicolumn{1}{|c|}{\multirow{2}{*}{6}} & \multicolumn{1}{|c|}{\multirow{2}{*}{20}} & [$1140$,$1860$] & ($1860$,$1980$) & [$1980$,$2700$]\\
&&  [$2160$,$2820$] &  ($2820$,$2860$) & [$2860$,$3600$]\\
\hline
\end{tabular}}
\end{center}
\end{table}

	In Table \ref{results}, the results for all tested scenarios are presented. Note that for the four first tests the Web application starts with two \textit{m3.large} vms. For the last two tests, the Web application starts with two \textit{m3.medium} vms, because the \textit{m3.medium} vm is enough to store all the contents.
	
	In all tested scenarios, the execution times were very similar for both approaches, meaning that no request attendance was delayed. Moreover, FCD mechanism takes, in the worst case, only $0,0003$s to calculate the values of the entropies and total correlation for each period of time, proving that it is efficient and does not add significant overhead in the proposed approach.

\begin{table}[htp]
\small
\tbl{Results for the tested scenarios.\label{results}}{
\begin{tabular}{|c|c|c|c|}
\hline 
Scenario &  FCD+FCHP-ILS &  AS+LB\\
\hline
\multicolumn{1}{|c|}{\multirow{2}{*}{1}} & 2 \textit{m3.large} + 6 \textit{m3.medium}	& 8 \textit{m3.large}	\\
	& \$12.13	 & \$13.62	\\
\hline
\multicolumn{1}{|c|}{\multirow{2}{*}{2}} & 2 \textit{m3.large} + 8 \textit{m3.medium}	& 10 \textit{m3.large}	\\
	& \$17.86  & \$19.28	\\
\hline
\multicolumn{1}{|c|}{\multirow{2}{*}{3}} & 2 \textit{m3.large} + 2 \textit{m3.medium}	& 4 \textit{m3.large}	\\
	& \$10.32 & \$11.39	\\
\hline
\multicolumn{1}{|c|}{\multirow{2}{*}{4}} & 2 \textit{m3.large} + 7 \textit{m3.medium}	& 9 \textit{m3.large}	\\
	& \$23.39 & \$25.46 	\\	
\hline
\multicolumn{1}{|c|}{\multirow{2}{*}{5}} & 9 \textit{m3.medium}	& 9 \textit{m3.medium}	\\
	& \$7.08 & \$8.18   \\
\hline
\multicolumn{1}{|c|}{\multirow{2}{*}{6}} & 7 \textit{m3.medium}	&	7 \textit{m3.medium} \\
	& \$6.21	 & \$6.77	\\
\hline

\end{tabular}}

\end{table}

	As can be seen in Table \ref{results}, in all scenarios our approach presents satisfactory results, reducing the financial costs. This is achieved because in our approach the new allocated virtual machines can be different from the originally allocated ones. It allows for the Web application to use heterogeneous resources of the clouds. Thus, cheaper and smaller virtual machines were allocated to store only the contents that were involved in the flash-crowd event.  
	Remark that, differently from our approach, the solution provided by Auto Scaling allocates virtual machines based on a static virtual image that must contain all contents of the Web application. So, if the Web application starts using \textit{m3.large} virtual machines, only this type of machine can be hired in the future.
	However, even in the last two scenarios, where both approaches start with the same type of virtual machine, our solution is better. This fact is justified by the cost of the use of the Load Balancing mechanism. In our approach, the frontend server, responsible for directing the received requests, can run in one of the started virtual machines of the Web application. Thus, no extra financial cost is added because this machine is already hired and does not exist communication cost between virtual machines. 

%% file: conclusion.tex
\section{Concluding remarks}
\label{sec:conc}

	In this paper, a new mechanism to detect flash crowd, named Flash-Crowd Detection (FCD), which uses the concepts of entropy and total correlation, is proposed. This mechanism was evaluated using different traces of real and synthetic flash-crowd events.
	
	Moreover, the Flash-Crowd Handling Problem (FCHP) is precisely defined and formulated as an integer programming problem. A new algorithm for solving it, named FCHP-ILS, is also proposed. With FCHP-ILS the Web provider is able to replicate contents in the available resources and define the types and amount of resources to instantiate in the cloud during a flash-crowd event. We evaluate FCHP-ILS, in terms of  quality of solution and execution time,  by comparing it with the solutions given by the formulation FCHP-IP. The results showed that the FCHP-ILS obtains good solutions with tight gaps, in average $1.8\%$, and it takes significantly less time, in average $18.5$s against $2466.9$s.
	
	Finally, the FCHP-ILS was combined with the FCD mechanism in order to provide detection and handling in a single approach. In order to analyze the performance of our approach in a  commercial cloud, we executed it in small real scenarios of the problem and compared with the mechanism provided by the popular Auto Scaling and Load Balancing of the Amazon. The results showed that the proposed approach for detecting and handling flash crowds is efficient when tested in a commercial cloud (Amazon) against the Auto Scaling method, reducing the  amount of instantiated resources on the cloud and, consequently, the financial costs, during flash crowds.

%% file: appendix.tex
\appendix
\section*{APPENDIX}\label{appendix}

In order to describe formulation FCHP-IP, we need to extend the notation defined in Section \ref{sec:model}. Let $S=S_p \cup S_e$ be the set of all servers (virtual or physical), where $S_p$ and $S_e$ are defined as the set of Web application servers and the set of servers available for hire in the cloud, respectively. We also consider a set of requests $R$ to be attended and a set of  contents $C$ offered by the Web application in a period of time $t\in T = \{T_1,\dots,T_{f}\}$. Each request $i \in R$ requires a content and each content $k \in C$  has a size $L_k$. Each server $j \in S$ has a storage capacity $AS_{j}$ and a maximal bandwidth $MB_{j}$ and each server $j \in S_e$ also has a financial cost $f_j$. Similarly, each content $k \in C$ has a start time $B_{k}$ (\textit{i.e.} the period that content $k$ is submitted in the application) and an origin server $O_{k}$. Moreover, each request $i \in R$ requires a content $g(i)$ and we denote as $D_{it}$, the demand of request $i \in R$ in period $t \in T$. Finally, we define $BX$ as the average bandwidth of the clients. 

Now we can restate the FCHP as the problem of copying replicas of contents on the servers and hiring new cloud servers in order to handle the requests during the flash crowd, respecting the available storage and bandwidth and trying to minimize the cost function defined by four sums of time: (i) the time cost $c_{i}$ to handle request $i \in R$, defined by $\sum_{i\in R}  \sum_{j \in S} \sum_{t\in T} c_{i}x_{ijt}$, where the binary variable $x_{ijt}$ indicates if the request $i$ is attended by server $j$ in period $t$, (ii) the sum of backlogging time penalty $p_{it}$ (\textit{i.e.} the penalty to postpone the attendance of request $i$ that arrived in period $t$), defined by $\sum_{i \in R} \sum_{t \in T} p_{it}b_{it}$, where $b_{it}$ indicates the postponed amount of request $i$ in period $t$, (iii) the sum of time cost $h_{k}$ to copy the content $k$, defined by $\sum_{k \in C} \sum_{j \in S} \sum_{l \in S} \sum_{t \in T} h_{k}w_{kjlt}$, where the binary variable $w_{kjlt}$ indicates if the content $k$ is copied from server $j$ to server $l$ in period $t$, and (iv) the sum of financial cost $f_j$ to hire the server $j$ in period $t$, defined by $ \sum_{j \in S} \sum_{t \in T} f_{j}z_{jt} \frac{1}{M} $, where the binary variable $z_{jt}$ indicates if the server $j$ is hired in period $t$ and $M=\max\{c_i,p_{it}b_{it},h_k,f_j\}$ $\forall i \in R, t \in T, k \in C $ and $j \in S_e$. Note that the fourth sum can not be greater than any other cost in the previous sums. So, this last sum works as a tiebreaker for solutions with the same time cost to obtain the one with the minimal financial cost. 
Furthermore, we also define other three binary variables: $y_{kjt}$ that indicates if content $k \in C$ is replicated on server $j \in S$ in period $t \in T$; $s_{iojt}$ that indicates if the demand of request $i \in R$ of period $o \in T$ (i.e. $D_{io}$) is attended by server $j \in S$ in period $t \in T$. 
Note that the expression $ \sum_{o=1}^{t} ( s_{iojt}D_{io} )$ is equivalent to the amount of demand attended of request $i$ in server $j$ on period $t$.
Finally, variable $z_{jt}$ indicates if server $j \in S_e$ is hired in period $\alpha \in \overline{T}=\{\overline{T_1},\ldots,\overline{T_f}\}$. Note that the time scale $\overline{T}$ can have a granularity greater than $T$ due to the minimal hiring period offer by the cloud provider. 
We also define two constants, $tr$ and $tp$, representing the number of periods that a replication takes to be available and the number of periods that a hired server takes to be available, respectively.
The FCHP can be formulated as the mixed integer programming problem, named FCHP-IP, described next.

\begin{align}
\mathrm{minimize\ } &  \sum_{i\in R}  \sum_{j \in S} \sum_{t\in T} c_{i}x_{ijt} + \sum_{r \in R} \sum_{t \in T} p_{it}b_{it} + \sum_{k \in C} \sum_{j \in S} \sum_{l \in S}\sum_{t \in T} h_{k}w_{kjlt} + \sum_{j \in S} \sum_{t \in T} f_{j}z_{jt} \frac{1}{M}& \label{fo4}
\end{align}
\begin{align}
\mathrm{subject\ to\ } & &\nonumber\\
                 & \sum_{j \in S} \Big(  \sum_{o=1}^{t} ( s_{iojt}D_{io} ) \Big)  = D_{it}+ (b_{i(t-1)}-b_{it}), & \forall i \in R,  \forall t \in [B_{g(i)},T_f] \label{r1}\\
                 & \sum_{i \in R} \Big(  \sum_{o=1}^{t} ( s_{iojt}D_{io} ) \Big) \leqslant MB_{j},            & \forall j \in S,  \forall t \in T  \label{r2}\\
                 & \sum_{j \in S} \Big(  \sum_{o=1}^{t} ( s_{iojt}D_{io} ) \Big) \leqslant BX,                & \forall i \in R, \forall t \in T  \label{r3}\\
                 & \sum_{j \in S} \sum_{t \in T}\Big(  \sum_{o=1}^{t} ( s_{iojt}D_{io} ) \Big) = L_{g(i)},    & \forall i \in R  \label{r4}\\
                 & \sum_{o=1}^{t} ( s_{iojt}D_{io} ) \leqslant x_{ijt}L_{g(i)},                               & \forall i \in R, \forall j \in S, \forall t \in T \label{r4-1}\\
                 & y_{g(i)jt} \geqslant x_{ijt},                                                              & \forall i \in R, \forall j \in S, \forall t \in T  \label{r5}\\
                 & y_{kO_{k}B_{k}} = 1,                                                                       &\forall k \in C \label{r6}\\
                 & \sum_{\substack{t=1\ldots\\(B_k)}} \sum_{j \in S} y_{kjt} = 0 ,                            &\forall k \in C \label{r7}\\
                 & \sum_{\substack{j \in S\\j \neq O_k}} y_{kjB_k} = 0 ,                                      & \forall k \in C \label{r8}\\
                 & \sum_{\substack{t=1\ldots\\(B_k)}} w_{kjlt} = 0 ,                                          & \forall k \in C, \forall j \in S, \forall l \in S \label{r9}\\
                 & y_{kj(t+tr)} \leqslant \sum_{l \in S}w_{kjlt},                                             & \forall k \in C, \forall j \in S,   \forall t \in [B_k,T_f] \label{r10} \\ 
                 & y_{kjt} \geqslant w_{kljt},                                                                & \forall k \in C, \forall j \in S, \forall l \in S,  \forall t \in [B_k,T_f]\label{r11}\\ 
                 & \sum_{k \in C} L_{k}y_{kjt} \leqslant AS_{j},                                              & \forall j \in S, \forall t \in T\label{r12}
\end{align}
\begin{align}             
                 & x_{ijt} \leq z_{j\alpha} ,                                                                 & \forall i\in R, \forall j \in S_e, \forall (t-tp) \in [\alpha-1,\alpha], \alpha \in \overline{T}\label{r13}\\
                 & x_{ijt} \in \{0,1\},       																& \forall i \in R, \forall j \in S,  \forall t \in T\label{r14-1}\\
                 & b_{it} \geqslant 0,																        & \forall i \in R, \forall j \in S,  \forall t \in T\label{r14-2}\\
                 & s_{iojt} \in \{0,1\},     															    & \forall i \in R, \forall j \in S,  \forall t,o \in T\label{r14-3}\\
                 & y_{kjt} \in \{0,1\},									                                    & \forall k \in C, \forall j \in S,  \forall t \in T\label{r15}\\
                 & z_{jt} \in \{0,1\},									                                    & \forall k \in C, \forall j \in S,  \forall t \in T\label{r15}\\
                 & w_{ijlt} \in \{0,1\},                                                                      &\forall j,l \in S,  \forall k \in C, \forall t \in T\label{r17}
\end{align}

%

The objective function (\ref{fo4}) minimizes the operational time costs, backlog and financial cost. 
Constraints (\ref{r1}) relate the demand attended of request $i \in R$ in period $t \in [B_{g(i)},T_f]$ (i.e. left-hand side of the equation) with the postponed and inherited backlog (i.e. $b_{it}$ and $b_{i(t-1)}$ respectively). 
Servers bandwidth are controlled by constraints (\ref{r2}). Constraints (\ref{r3}) prevent that a client receive more than it can bear. Constraints (\ref{r4}) guarantee that every request is fully handled. 
Constraints (\ref{r4-1}) relate the variables $s_{iojt}$ and $x_{ijt}$. Note that, if $s_{iojt}=1$ for any $o \in T$, then we should impose $x_{ijt}=1$.
Constraints (\ref{r5}) rule that a request must be handled by a server that has a replica of the desired content. Constraints (\ref{r6}), (\ref{r7}), (\ref{r8}) and (\ref{r9}) control the number of replicas of active contents and make sure that only the origin server has a replica of the content on the submission period. Constraints (\ref{r10}) guarantee that all replications create a new replica. Constraints (\ref{r11}) make sure that a replication occurs from a server that has the content. The servers storage capacity are controlled by constraints (\ref{r12}). Constraints (\ref{r13}) impose that a request must be handled by a hired server. The remaining constraints are the integrality and non-negativity constraints.